\documentclass[floatfix,aps,prd,abstract,twocolumn,10pt,nofootinbib,preprintnumbers]{revtex4}

\usepackage{amsgen,amsmath,amstext,amsbsy,amsopn,amssymb}
\usepackage{graphicx}
\usepackage[usenames]{color}
\usepackage{epsfig}
\usepackage{longtable}
\usepackage{bm} 
\usepackage{ulem}
\usepackage{hyperref}
\usepackage{float}

\newcommand{\Ddel}{\delta_{\rm D}   }

\newcommand{\MpcOh}{ \,  \mathrm{Mpc}  \, h^{-1} }
\newcommand{\hOMpc}{ \,  \mathrm{Mpc}^{-1}  \, h  }

\newcommand{\comment}[1]{}

\newcommand{\beq}{\begin{equation}}
\newcommand{\eeq}{\end{equation}}
\newcommand{\beqa}{\begin{eqnarray}}
\newcommand{\eeqa}{\end{eqnarray}}
\newcommand{\nn}{ \nonumber }

\newcommand{\cyc}{\, \mathrm{cyc.} }
\allowdisplaybreaks[1]

\begin{document}


\title{ Bispectrum Supersample Covariance }

\author{Kwan Chuen Chan} \email{chankc@mail.sysu.edu.cn}
\affiliation{ School of Physics and Astronomy, Sun Yat-Sen University, Guangzhou 510275, China} 
\affiliation{ Institut d'Estudis Espacials de Catalunya (IEEC), 08193 Barcelona, Spain }
\affiliation{ Institute of Space Sciences (ICE, CSIC),  Campus UAB, s/n,  08193 Barcelona, Spain }

\author{Azadeh Moradinezhad Dizgah}
\affiliation{Department of Physics, Harvard University, 17 Oxford Street, Cambridge, MA 02138, USA }
\author{Jorge Nore\~na}
\affiliation{Instituto de F\'{i}sica, Pontificia Universidad Cat\'{o}lica de Valpara\'{i}so, Casilla 4059, Valpara\'{i}so, Chile}

\begin{abstract}
  Modes with wavelengths larger than the survey window can have significant impact on the covariance within the survey window.  The supersample covariance has been recognized as an important source of covariance for the power spectrum on small scales, and it can potentially be important for the bispectrum covariance as well. In this paper, using the response function formalism, we model the supersample covariance contributions to the bispectrum covariance and the cross covariance between the power spectrum and the bispectrum. The supersample covariances due to the long wavelength density and tidal perturbations are investigated, and the tidal contribution is a few orders of magnitude smaller than the density one because in configuration space the bispectrum estimator involves angular averaging and the tidal response function is anisotropic.  The impact of the super-survey modes is quantified using numerical measurements with periodic box and subbox setups. For the matter bispectrum, the ratio between the supersample covariance correction and the small scale covariance, which can be computed using a periodic box,  is roughly an order of magnitude smaller than that for the matter power spectrum.  This is because for the bispectrum, the small scale non-Gaussian covariance is significantly larger than that for the power spectrum. For the cross covariance, the supersample covariance is as important as for the power spectrum covariance.  The supersample covariance prediction with the halo model response function is in good agreement with numerical results.

\end{abstract}

\maketitle

\section{ Introduction } 

Current photometric and spectroscopic large scale structure surveys, such as DES \cite{Abbott:2017wau,Abbott:2017wcz} and BOSS \cite{Alam:2016hwk}, have contributed significantly in improving our understanding of the early- and late-time universe. This trend will continue in the future through upcoming surveys such as Euclid \cite{EuclidRedBook} and LSST \cite{LSSTScienceBook} as they are expected to cover a larger volume and wider redshift range with an unprecedented precision.
The key question is how much cosmological information can be extracted from such high-fidelity data.  In particular, we are interested in quantifying the information content of higher-order correlation functions, focusing on the bispectrum. An important ingredient required to answer this question is to correctly model the covariance matrix.

The covariance of the polyspectra can be generally classified into three parts: the Gaussian covariance due to the random phases of the Fourier modes, the non-Gaussian covariance due to the mode coupling between the modes inside the survey window (we sometimes call this the small-scale covariance), and the covariance due to the coupling of the modes outside the survey window with those inside.  The small-scale covariance can be studied using the standard periodic boundary condition setup. 
Ref.~\cite{Hamilton:2005dx} pointed out that because of the presence of the window function in a real survey, the long modes larger than the survey window size can modulate the small scale modes and lead to large covariance on small scales.  The authors coined the term beat coupling to refer to the covariance due to the long mode outside the window.   The wave vectors are sharp in simulations with periodic boundary conditions, and so this type of covariance cannot be studied in the standard periodic box setup; instead it can be studied by dividing a gigantic simulation box into multiple subboxes.   Ref.~\cite{Takada:2013bfn}  formulated this type of covariance using the response function formalism. In this work we follow \cite{Takada:2013bfn} in referring to the covariance due to the modes outside the survey window as the supersample covariance and the perturbative part of it as beat coupling since these terminologies are widely used now.  The response function approach borrows the technique of the consistency relation, first derived in the context of inflation \cite{Maldacena:2002vr,Creminelli:2004yq}, and later applied in large scale structure context \cite{Peloso:2013zw, Kehagias:2013yd, Creminelli:2013mca, Horn:2014rta}. The response approach provides a powerful scheme to model the coupling of the long mode with the small scale modes.

In previous studies on the information content of the bispectrum, a Gaussian covariance was assumed, e.g.~\cite{Takada:2003ef,Sefusatti:2004xz}.  In the context of weak lensing, Refs.~\cite{Kayo:2013aha,KayoTakadaJain_2013,Sato:2013mq} found that when using realistic non-Gaussian covariance, the information content of the lensing bispectrum is overestimated relative to the Gaussian covariance approximation.  Recently, \cite{Chan:2016ehg} studied the bispectrum covariance matrix using a large suite of simulations, and found that the Gaussian covariance significantly overestimates the information content since the Gaussian covariance approximation is a poor approximation beyond the mildly linear regime (see \cite{Sefusatti:2006pa,Byun:2017fkz} for the constraint on the cosmological parameters).  However, \cite{Chan:2016ehg} measured the covariance from periodic simulations, and so the supersample covariance was  not present. It is the goal of this paper to address how important the supersample covariance is to the budget of the bispectrum covariance.

This paper is organized as follows.  In Sec.~\ref{sec:bisp_ssc_derivation}, using the response function formalism, we derive the supersample bispectrum covariance and  the cross covariance between the power spectrum and the bispectrum. The general bispectrum response to the long mode is studied in Sec.~\ref{sec:Bk_response_effect_general}.  We compute the bispectrum response function using the standard perturbation theory and halo model in Sec.~\ref{sec:Bk_reponse_SPT_HM}.  In Sec.~\ref{sec:predictions_measurements}, we quantify the magnitudes of the supersample covariance on the bispectrum covariance and the cross covariance by comparing the numerical measurements obtained from the periodic box and subbox setups. We also compare the predictions obtained with the halo model prescription  with the numerical results. We conclude in Sec.~\ref{sec:conclusions}. In Appendix ~\ref{sec:beat_coupling_derivation}, we compute the supersample covariance using a simple beat coupling approach and check it against the response formalism.  We generalize the calculations to compute the effect of the tidal perturbations on the bispectrum supersample covariances in Appendix ~\ref{sec:SSC_tides}.

\section{Derivation of the bispectrum supersample covariance}
\label{sec:bisp_ssc_derivation}

In this section, we derive the supersample covariance contributions to the bispectrum covariance and the cross covariance between the power spectrum and the bispectrum. We model the supersample covariance using the response function formalism.  The derivation is a straightforward generalization of the computation of the supersample covariance for the power spectrum in Ref. \cite{Takada:2013bfn}.

\subsection{ The bispectrum estimator and window function }

Suppose we have measured the Fourier mode of the density contrast, $\hat{\delta}(\bm{k})$, from a survey or simulation.   To estimate the bispectrum, the Fourier modes are binned into shells of width $\Delta k $.  From the definition of the bispectrum
\beq
 \langle   \hat{\delta}(\bm{k}_1)  \hat{\delta}(\bm{k}_2)  \hat{\delta}(\bm{k}_3)  \rangle = ( 2 \pi)^3 \Ddel( \bm{k}_{123} ) \hat{B}(k_1,k_2,k_3) ,
 \eeq
(where $  \Ddel $ is the Dirac delta function and $\bm{k}_{123}$ denotes $\bm k_1 + \bm k_2 + \bm k_3 $)  we can construct an estimator as \cite{SCFFHM98,Scoccimarro:2003wn} 
\begin{align}
\label{eq:Bisp_estimator}
  \hat{B}(k_1,k_2,k_3)  
& =  \frac{ 1 }{  V  V_{\triangle} }\int_{ k_1 }d^3 p \int_{ k_2 }d^3 q \int_{ k_3 }d^3 r  \,   \  \nn \\  
& \times   \Ddel( \bm{p} + \bm{q} + \bm{r} ) \,  \hat{\delta}(\bm{p})  \hat{\delta}(\bm{q} )  \hat{\delta}(\bm{r} ) , 
\end{align}
where $k_i$ indicates that the integration is over a spherical shell of width $[ k_i - \Delta k /2 , k_i + \Delta k /2 ) $. $V$ is the volume of the survey/simulation, and  $ V_{\triangle }$ counts the number of modes satisfying the triangle condition 
\beq
\label{eq:Vtriangle}
V_{ \triangle }(k_1,k_2,k_3) = \int_{k_1} d^3 p \int_{k_2} d^3 q \int_{k_3} d^3 r \,   \Ddel( \bm{p} + \bm{q} + \bm{r} ).
\eeq 
$ V_{ \triangle } $ can be computed analytically (\cite{Scoccimarro:2003wn}, see \cite{Chan:2016ehg} for a review of the derivation) 
\beq
\label{eq:Vtriangle_final}
V_{\Delta} =   8 \pi^2 k_1 k_2 k_3 ( \Delta k )^3 \beta(\mu ) , 
\eeq
where $\mu  $ is defined as 
\beq
\label{eq:mu_triangle}
\mu = \hat{\bm{k}}_1 \cdot \hat{\bm{k}}_2 =  \frac{ k_3^2 - k_1^2 -  k_2^2 }{ 2 k_1 k_2  },
\eeq
and $\beta (\mu) $ is given by
\beq
\label{eq:beta_mu}
\beta(\mu )  = \begin{cases} \frac{1}{2}  \, & \mbox{if } \mu  =  \pm 1    \\
                                1               &\mbox{if }  0  < \mu < 1   \\
                                0               &\mbox{otherwise }  
              \end{cases} .
\eeq

In a realistic survey, there is a survey window function and the measured $\hat{ \delta}(\bm{k})$  is a convolution of the survey window with the underlying density field. Here we study the implications of the survey window on $\hat{B}$.

The  survey volume $V$ can be expressed in terms of a general window function $W$ \footnote{ The survey window function considered here is dimensionless in real space, and its value in real space falls within  the interval [0,1]. This is different from the one used to define Lagrangian halos (e.g., \cite{Chan:2015zjt}), which has the dimension of inverse volume.  In that case, the window function convolves with the density field in real space, and the size of the window is close to the Lagrangian size of the halo. },
 \beq
V = \int d^3 x W(\bm{x} ). 
\eeq
The density contrast in real space $ \delta_{W} (\bm{x}  ) $ reads
\beq
\delta_{W} (\bm{x}  )  = W(\bm{x} ) \delta(\bm{x}),
\eeq
where  $ \delta(\bm{x}) $ can be the density contrast of the dark matter or other tracers. In Fourier space, we have
\beq
\label{eq:deltaW}
\delta_W(\bm{k})  = \int \frac{d^3 p }{ (2 \pi)^3 } \delta(\bm{p}) W(\bm{k} - \bm{p} ).   
\eeq
Thus the effect of the selection window is to smooth the density contrast in Fourier space. The width of the window $W$ is of the order $1/L$, where $L= V^{1/3}$.  The wave vector is effectively broadened by $\sim 1/L$ in Fourier space. In contrast, for simulations with periodic boundary conditions, only wave vectors in units of the fundamental mode are supported and hence they are sharp.   This is why the window function effect is not captured by the standard periodic simulation setup.


Plugging the smoothed density Eq.~\eqref{eq:deltaW} into the estimator $\hat{B}$, we get
\begin{align}
\label{eq:B_w_form1} 
  \hat{B}_W ( k_1, k_2,k_3) 
 &= \frac{ 1 }{V V_{\triangle} } \int_{k_1} d^3 p_1 \int_{k_2} d^3 p_2 \int_{k_3} d^3 p_3 \Ddel(\bm{p}_{123} )   \nn \\
&  \quad  \times   \prod_{i=1}^3 \int \frac{d^3 q_i }{(2 \pi)^3  }  W(\bm{q}_i ) \delta( \bm{p}_i - \bm{q}_i ). 
\end{align}
We are interested in how the broadening of the Fourier modes affects the estimator.  Because $ q_i \lesssim 1/L $, to extract the effect of the long mode $ q_i$  we take the limit $ q_i \ll k_j$. In this limit, we can do a change of variables without modifying the integration limits of $\bm{p}$-integrals, and write Eq.~\eqref{eq:B_w_form1} as
\begin{align}
\label{eq:B_w_form2} 
&  \hat{B}_W ( k_1, k_2,k_3) 
\approx  \frac{ 1 }{V V_{\triangle} } \int_{k_1} d^3 p_1 \int_{k_2} d^3 p_2 \int_{k_3} d^3 p_3  \Ddel(\bm{p}_{123} )  \nn \\
&  \quad \quad  \times   \prod_{i=1}^3 \int \frac{d^3 q_i }{(2 \pi)^3  }  W(\bm{q}_i ) \delta( \bm{p}_1 )  \delta( \bm{p}_2 )  \delta( \bm{p}_3 - \bm{q}_{123} ).   
\end{align}
Eq.~\eqref{eq:B_w_form2} reveals that although we try to measure the bispectrum of  $ \delta_W $ satisfying the closed triangle condition by imposing the Dirac delta function in Eq.~\eqref{eq:Bisp_estimator},  the presence of the window function opens the triangle slightly by an amount $\bm{q}_{123} $  for $ \delta $.  Analogously, for the case of the power spectrum, although we try to measure the power spectrum of $ \delta_W $ using wave vectors that are equal in magnitude but opposite in direction, the long mode causes a slight misalignment of these two vectors for $ \delta $.

In Eq.~\eqref{eq:B_w_form2}, we have isolated the effect of the long mode in one of the Fourier modes to facilitate the analysis later on. This form appears to break the symmetry among $k_1$, $k_2$, and $k_3$; however, this breaking is of higher order in $q_i$  and our final results will be symmetric about  $k_1$, $k_2$, and $k_3$. Our only approximation in Eq.~\eqref{eq:B_w_form2} is that the limits of $\bm{p}$-integrals are unchanged. The effect is expected to be small as it only slightly changes the total number of configurations satisfying the constraint, while the dominant effect comes from the fact that \textit{each} of the triangle configurations is opened by the long mode.

\begin{widetext}

\subsection{ Effect of the long mode on the covariance } 
\label{sec:SSC_Bk_derivation}

The window function convolves the bispectrum in Fourier space, and hence it can bias the amplitude and imprint wiggles on the measured bispectrum. We will discuss this more later on.  In this section we are interested in the effect of the long mode on the small scale measurements. To leading order, the effect of the long mode on the expectation value of $ \hat B_W $ vanishes.

We now examine how the window function affects the covariance of the estimator $\hat{B}_W$. The covariance of $\hat{B}_W$ is given by
\beq
\mathrm{cov} \big( \hat{B}_W(k_1, k_2,k_3),  \hat{B}_W(k_1', k_2',k_3')  \big) = \langle \hat{B}_W (k_1, k_2,k_3 )  \hat{B}_W(k_1',k_2', k_3')  \rangle - \langle \hat{B}_W (k_1, k_2,k_3 ) \rangle  \langle \hat{B}_W(k_1',k_2', k_3')  \rangle.
\eeq

Our task is to compute the connected part of $\langle \hat{B}_W (k_1, k_2,k_3 )  \hat{B}_W(k_1',k_2', k_3')   \rangle   $ due to the long mode
\begin{align}
  \label{eq:BB_expectation}
   \langle  \hat{B}_W (k_1, k_2,k_3 ) \hat{B}_W(k_1',k_2', k_3')   \rangle  
   =& \frac{1  }{ V^2  V_{\triangle} V_{\triangle}' } \prod_{j=1}^3 \int_{k_j} d^3 p_j  \Ddel( \bm{p}_{123})   \int_{k_j'} d^3 p_j'  \Ddel( \bm{p}_{123}')   \prod_{i=1}^3 \int \frac{d^3 q_i}{(2 \pi)^3 } W(q_i) \int \frac{d^3 q_i'}{(2 \pi)^3 } W(q_i')                \nn \\
  &\qquad  \times \langle  \delta(\bm{p}_1)   \delta(\bm{p}_2)   \delta(\bm{p}_3 - \bm{q}_{123} )   \delta(\bm{p}_1')  \delta(\bm{p}_2')  \delta(\bm{p}_3' - \bm{q}_{123}' )    \rangle.
\end{align}
The effect of the long modes $\bm{q}_{123}$ and  $\bm{q}'_{123}$ on the small scale bispectrum can be computed similar to \cite{Takada:2013bfn} by employing the argument of consistency relations for a soft internal mode
\begin{align}
  \label{eq:6point_longmode_expand1} 
   \Big\langle  \delta(\bm{p}_1)   \delta(\bm{p}_2)   \delta(\bm{p}_3 - \bm{q}_{123} )  &  \delta(\bm{p}_1')  \delta(\bm{p}_2')  \delta(\bm{p}_3' - \bm{q}_{123}' )    \Big\rangle 
  \approx  \Big\langle  \Big[ \langle  \delta(\bm{p}_1)   \delta(\bm{p}_2)   \delta(\bm{p}_3) \rangle  + \delta_{\rm l}( - \bm{q}_{123} ) \frac{ \partial  }{ \partial \delta_{\rm l}( \bm{q} )   }   \langle  \delta(\bm{p}_1)   \delta(\bm{p}_2)   \delta(\bm{p}_3 + \bm{q}) \rangle \Big|_{\delta_{\rm l} =0 } \Big]    \nn \\
 & \qquad  \times    \Big[ \langle  \delta(\bm{p}_1')   \delta(\bm{p}_2')   \delta(\bm{p}_3') \rangle  + \delta_{\rm l} (- \bm{q}_{123}' ) \frac{ \partial  }{ \partial \delta_{\rm l} (\bm{q}')  } \langle  \delta(\bm{p}_1')   \delta(\bm{p}_2')   \delta(\bm{p}_3' + \bm{q}' ) \rangle \Big|_{\delta_{\rm l} =0 } \Big]   \Big\rangle_{\delta_{\rm l}} , 
\end{align}
where the expectation value sign $\langle \dots \rangle_{\delta_{\rm l} } $ denotes the average over the long mode ${\delta_{\rm l} } $, while inside the expectation value sign $\delta_{\rm l} $ is kept fixed.  The long wavelength perturbation can be expressed as
\beq
\label{eq:delta_b_3Dform}
\delta_{\rm l} ( \bm{q} ) = (2 \pi)^3 \Ddel( \bm{ q} ) \delta_{\rm b} ,
\eeq
where $ \delta_{\rm b} $ is the \textit{dimensionless} amplitude of perturbation. Then Eq.~\eqref{eq:6point_longmode_expand1} can be written as 
\begin{align}
  \label{eq:6point_longmode_expand2} 
  & \Big\langle  \delta(\bm{p}_1)   \delta(\bm{p}_2)   \delta(\bm{p}_3 - \bm{q}_{123} )   \delta(\bm{p}_1')  \delta(\bm{p}_2')  \delta(\bm{p}_3' - \bm{q}_{123}' )    \Big\rangle \nn \\
  = &  \langle  \delta(\bm{p}_1)   \delta(\bm{p}_2)   \delta(\bm{p}_3 ) \rangle  \langle \delta(\bm{p}_1')  \delta(\bm{p}_2')  \delta(\bm{p}_3' )   \rangle    +   \langle  \delta_{\rm l}( - \bm{q}_{123} )   \delta_{\rm l}( - \bm{q}_{123}' )  \rangle  \frac{\partial}{\partial \delta_{\rm b} } B(p_1,p_2,p_3 | \delta_{\rm b} )   \Big|_{\delta_{\rm b} = 0 }    \frac{\partial}{\partial \delta_{\rm b} }   B(p_1',p_2',p_3'|\delta_{\rm b})  \Big|_{\delta_{\rm b} = 0 }  \nn \\
  =&  \langle  \delta(\bm{p}_1)   \delta(\bm{p}_2)   \delta(\bm{p}_3 ) \rangle  \langle \delta(\bm{p}_1')  \delta(\bm{p}_2')  \delta(\bm{p}_3' )   \rangle  + (2 \pi)^3  P_{\rm l}(q_{123} ) \Ddel( \bm{q}_{123} + \bm{q}'_{123} ) \frac{\partial}{\partial \delta_{\rm b} } B(p_1,p_2,p_3 | \delta_{\rm b}   )   \Big|_{\delta_{\rm b} = 0 }    \frac{\partial}{\partial \delta_{\rm b} }   B(p_1',p_2',p_3' | \delta_{\rm b}  )  \Big|_{\delta_{\rm b} = 0 }  . 
\end{align}
The first term in Eq.~\eqref{eq:6point_longmode_expand2} is canceled by $\langle B_W \rangle  \langle B_W' \rangle $, and only the second one  contributes to the bispectrum covariance.  Note that in Eq.~\eqref{eq:6point_longmode_expand2}, $P_{\rm l } $ is the power spectrum of the long mode and it is assumed to be linear, while the bispectrum $B$ can be highly nonlinear.

For the power spectrum covariance, an analogous relation, which was called the trispectrum consistency relation in Ref. \cite{Takada:2013bfn}, can be established.  The consideration of the effects of the long mode on short scales is the key to construct the large-scale structure  consistency relation \cite{Peloso:2013zw, Kehagias:2013yd, Creminelli:2013mca, Horn:2014rta}.  The position-dependent power spectrum, which is equivalent to squeezed bispectrum, is constructed  by isolating the effects of the long modes on the local power spectrum \cite{Chiang:2014oga} (see \cite{Adhikari:2016wpj} for a generalization to the position-dependent bispectrum).

Plugging the second term in the last line of Eq.~\eqref{eq:6point_longmode_expand2} into Eq.~\eqref{eq:BB_expectation}, we can perform the $\bm{q} $ and $\bm{q}'$ integrals as
\begin{align}
  \label{eq:Wconvolution_intg}
 & \frac{1}{V^2} \prod_{i=1}^3 \int \frac{d^3 q_i}{(2 \pi)^3 } W(q_i) \int \frac{d^3 q_i'}{(2 \pi)^3 } W(q_i')   (2 \pi)^3 \Ddel( \bm{q}_{123} +  \bm{q}'_{123} )  P_{\rm l} (q_{123} ) \nn \\ 
 = & \frac{1}{V^2} \prod_{i=1}^3 \int \frac{d^3 q_i}{(2 \pi)^3 } W( \bm{q}_i ) W_3( \bm{q}_{123}  )  P_{\rm l} (q_{123} ) \nn \\
 =&  \frac{1  }{ V^2 } \prod_{i=1}^3  \int \frac{ d^3 Q_i }{ (2\pi)^3 }  P_{\rm l}(Q_3) W_3( Q_3)   \int \frac{ d^3 Q_3' }{ (2\pi)^3 } (2 \pi)^3 \Ddel( \bm{Q}_3' - \bm{Q}_3  + \bm{ Q}_{12}  ) W(Q_1) W(Q_2) W( Q_3' ) \nn \\
 = &  \int \frac{ d^3 Q_3 }{(2\pi)^3 } \bigg[ \frac{ W_3 ( Q_3) }{V} \bigg]^2P_{\rm l} (Q_3) \equiv \sigma_{W_3}^2 .
\end{align}
In the first equality, we have simply defined the notation $W_n$ (with $n=3$)
\beq
W_n ( \bm{k} ) \equiv  \int \frac{d^3 k_1  }{ (2 \pi)^3 } \int \frac{d^3 k_2  }{ (2 \pi)^3 } \dots \int \frac{d^3 k_n  } { (2 \pi)^3 }  ( 2 \pi)^3 \Ddel( \bm{k} - \bm{k}_{12 \dots n } ) W( \bm{k}_1) \dots  W( \bm{k}_n), 
\eeq
and in real space we have  $W_n ( \bm{x} ) =  W^n ( \bm{x} )$.  For the second equality we have changed the variables as $\bm{Q}_1= \bm{q}_1 $,  $\bm{Q}_2= \bm{q}_{2} $, and  $\bm{Q}_3= \bm{q}_{123} $, and have explicitly introduced a Dirac delta function for $ \bm{Q}_3'$.  By including the volume in  the definition, $ \sigma_{W_3}^2 $ is the usual RMS variance of the long wavelength fluctuations across the survey window computed using $W_3$.

Finally we arrive at the supersample covariance for the bispectrum 
\beq
\label{eq:CB_SSC}
C^B_{\rm SSC}(k_1,k_2,k_3, k_1',k_2',k_3') =  \sigma_{W_3}^2  \frac{\partial }{ \partial \delta_{\rm b} } B( k_1,k_2,k_3 | \delta_{\rm b} ) \Big|_{\delta_{\rm b} = 0 } \frac{\partial }{ \partial \delta_{\rm b} } B( k_1',k_2',k_3' | \delta_{\rm b} ) \Big|_{\delta_{\rm b} = 0 }.  
\eeq
We call $\partial B / \partial \delta_b |_{ \delta_{\rm b} =0 } $ the bispectrum response function.

The supersample covariance for the power spectrum derived in \cite{Takada:2013bfn} is similar to Eq.~\eqref{eq:CB_SSC}, simply with $B$ replaced by $P$  and $ \sigma_{W_3}^2 $ replaced by $ \sigma_{W_2}^2 $, which is defined by substituting $W_3 $ with  $W_2 $ in the definition of  $ \sigma_{W_3}^2 $ [\cite{Takada:2013bfn} only explicitly considered the specific window Eq.~\eqref{eq:window_0_1}, so $W_2$ and $W_3$ are the same].

It is worth stressing that the perturbative expansion in Eq.~\eqref{eq:6point_longmode_expand2} is about the long wavelength mode that opens up the triangle, and it is distinctly different from the perturbative expansion about the small scales $\delta $ studied in \cite{Chan:2016ehg}. The supersample covariance arises from the coupling of the long mode with the small scale modes, while the non-Gaussianity investigated in  \cite{Chan:2016ehg} is purely from small scale couplings.

In Appendix \ref{sec:SSC_tides}, we extend the computations to include the supersample covariance contributions due to the tidal perturbations. The final result, Eq.~\eqref{eq:SSC_tide_appendix} is analogous to the density one Eq.~\eqref{eq:CB_SSC}.

\
\end{widetext}

\subsection{Supersample cross covariance between the power spectrum and the bispectrum}

As a by-product, it is straightforward to compute the supersample cross covariance between the power spectrum and the bispectrum. The power spectrum can be estimated by (e.g.~\cite{FeldmanKaiserPeacock1994,Scoccimarro:1999kp})
\beq
\label{eq:Pk_estimator}
\hat{P}(k) =   \frac{1}{V}   \int_{k}  \frac{ d^3 p }{ V_{\rm s}( k ) } \hat{\delta}( \bm{p} ) \hat{\delta}( - \bm{p} ), 
\eeq 
where the integration is over a spherical shell of width  $[ k - \Delta k /2,  k + \Delta k /2 ) $. $ V_{\rm s} $ is the volume of the spherical shell  
\beq
 V_{\rm s} (k) = \int_k d^3 p = 4 \pi k^2 \Delta k + \frac{\pi}{3} \Delta k^3. 
\eeq
The cross covariance between the power spectrum and the bispectrum is then given by
\beq
\mathrm{cov}( \hat{P}, \hat{B}  )= \langle  \hat{P} \hat{B} \rangle  -   \langle  \hat{P}\rangle \langle \hat{B} \rangle.
\eeq

Similar to the derivation in Sec.~\ref{sec:SSC_Bk_derivation}, it is easy to show that the supersample covariance contribution to cross covariance  is given by
\begin{align}
  \label{eq:SSC_PB_prediction}
   & C_{\rm SSC}^{PB}(k, k_1,k_2,k_3) \nn \\
  = & \sigma_{W_{2,3}}^2  \frac{ \partial  }{ \partial \delta_b } P(k| \delta_{\rm b} ) \bigg|_{ \delta_{\rm b} = 0 }   \frac{ \partial  }{ \partial \delta_b } B( k_1, k_2, k_3 | \delta_{\rm b} ) \bigg|_{ \delta_{\rm b} = 0 }  ,
\end{align}
where the mixed window variance $ \sigma_{W_{2,3}}^2 $ is defined as 
\beq
 \sigma_{W_{2,3}}^2 =   \int \frac{ d^3 Q }{(2\pi)^3 }  \frac{ W_2 ( Q ) }{V} \frac{ W_3 ( Q ) }{V}  P_{\rm l} (Q) .
\eeq

As the effects of a general window function is captured in the variance computed using $W_n $; now, similar to \cite{Takada:2013bfn}, we specialize to the window function $W$
\beq
\label{eq:window_0_1}
W(\bm{x} )  = \begin{cases} 1  \, & \mbox{inside survey }   \\
                                0               &\mbox{otherwise }  
              \end{cases} .
\eeq
This window function obeys the nice property that
\beq
\label{eq:Wn_identity_real}
W ( \bm{x} ) = W^n( \bm{x} ),  \quad  W ( \bm{k} ) = W_n( \bm{k} ).
\eeq
Hence, all of the variances are $\sigma_W^2  $ computed using $W$. For the rest of the paper, we will use the form Eq.~\eqref{eq:window_0_1} for $W$.

\section{ The response of the bispectrum to the long mode }
\label{sec:Bk_response_effect_general}

The long mode can affect the local measurement of the polyspectra in three ways: shifting the mean density used to define the density contrast, modifying the scale factor of the local patch, and  changing the intrinsic growth \cite{Hamilton:2005dx,Baldauf:2011bh,Sherwin:2012nh,dePutter:2012, Kehagias:2013paa,Takada:2013bfn,Li:2014sga, Wagner:2015gva, Baldauf:2015vio}.  These effects can be understood using the separate universe picture \cite{Sirko:2005uz,Baldauf:2011bh,Sherwin:2012nh,Li:2014sga,Wagner:2014aka,Baldauf:2015vio}, in which the long wavelength perturbation is absorbed into the background of a separate curved universe. We now describe each of them separately.

First, the density contrast is defined relative to the mean density, and we need to distinguish between the local and global mean densities \cite{dePutter:2012}. For galaxy surveys the density contrast is defined with respect to the local density contrast, while for weak lensing the global density is used  \cite{Takada:2013bfn}. The global and local mean densities, $\bar{\rho} $ and $\bar{\rho}_{W} $, are related by 
\beq
\rho(\bm{x}) = \bar{\rho} ( 1 + \delta(\bm{x}) ) =  \bar{\rho}_{W}( 1 + \delta_{W} (\bm{x}) ) ,   
\eeq
where $\delta$ and $\delta_{W}$ are the global and local density contrast. Because the global mean density and the local one are related by the background perturbation as 
\beq
\bar{\rho}_{W} = ( 1 + \delta_{\rm b} ) \bar{\rho} , 
\eeq
we have
\beq
\delta_{W }(\bm{x}  ) = \frac{\delta (\bm{x} ) - \delta_{\rm b}}{1 + \delta_{\rm b} }.  
\eeq
Or in Fourier space, we get
\beq
\delta_{W}(\bm{k})  = \frac{\delta (\bm{k}) - \delta_{\rm b}\Ddel(\bm{k})}{1 + \delta_{\rm b} }. 
\eeq
As we consider finite \textit{external} wave numbers, the Dirac delta function will not contribute, and so if the bispectrum is defined with respect to the local density we make the replacement
\beq
\label{eq:global_local_mean_replacement} 
B(k_1,k_2,k_3) \rightarrow \frac{ B(k_1,k_2,k_3) }{ ( 1 + \delta_{\rm b} )^3 }. 
\eeq

Second, the long mode modifies the background expansion rate in the local patch. By absorbing the long mode into the background density, the scale factor of the local universe, $a_W$  is related to the global one,  $a$ as  \cite{Sirko:2005uz,Baldauf:2011bh,Sherwin:2012nh,Li:2014sga,Wagner:2014aka}
\beq
\label{eq:density_pert_a3scaling}
\frac{ 1+ \delta_{\rm b}  }{ a^3 } = \frac{ 1}{ a_W^3 }. 
\eeq
The separate universe and the global universe describe the same physical system in different ways, thus the physical quantities in these descriptions must agree. By matching the physical length scale in these two universes, we infer that the comoving wave number in the local universe $\bm{k}_W$ is related to the global one $\bm{k}$ as \cite{Sherwin:2012nh,Li:2014sga,Wagner:2015gva}
\beq
\label{eq:kW}
\bm{k}_W = ( 1 + \delta_{\rm b} )^{ - \frac{1}{3} }  \bm{k}.  
\eeq
In Ref.~\cite{Li:2014sga}, this rescaling of the wave number was referred to as  the dilation effect.   As we shall see in Sec.~\ref{sec:response_SPT_longshortcoupling}, the dilation effect is incorporated into the standard perturbation theory.
Taking into account the transformation of the Dirac delta function, the dilation effect on the bispectrum is given by 
\beq
 B(k_1,k_2,k_3)  =  \frac{  B( k_{W1},  k_{W2},  k_{W3} ) }{ (1 - \delta_{\rm b} )  } . 
\eeq

The last effect is the modification of the intrinsic growth. If the background perturbation is positive, then gravity is stronger in the local universe, and so the intrinsic growth is enhanced. This effect can be studied by separate universe simulation \cite{Baldauf:2011bh,Li:2014sga,Wagner:2014aka}, perturbation theory \cite{Hamilton:2005dx}, or non-perturbative models such as hyper extended perturbation theory \cite{Hamilton:2005dx} or halo model \cite{Takada:2013bfn}. We compute this effect for dark matter bispectrum using perturbation theory in Sec.~\ref{sec:response_SPT_longshortcoupling} and the halo model in Sec.~\ref{sec:bisp_response_HM}.

  
We are now ready to check how the bispectrum response function  depends on these effects.  If the global mean density is used, the bispectrum response function is given by
\begin{align}
&   \frac{\partial  }{\partial \delta_{\rm b} }  \frac{  B_W  (k_{W1},  k_{W2},  k_{W3} ) }{ 1 - \delta_{\rm b}  }   \bigg|_{\delta_{\rm b}=0 } \nn \\ 
=  &   B( k_{1},  k_{2},  k_{3} )  + \frac{ \partial}{ \partial \delta_{\rm b}  } B_W( k_{W1},k_{W2},k_{W3}) \bigg|_{\delta_{\rm b}=0 } .
\end{align}
We use $B_W$ to denote the bispectrum resulting from the modified intrinsic growth.  As in \cite{Li:2014sga}, the second term can be analyzed by the chain rule
\begin{align}
  \label{eq:dBw_ddeltab_chainrule}
  & \frac{ \partial}{ \partial \delta_{\rm b}  }  B_W( k_{W1},k_{W2},k_{W3})  \bigg|_{\delta_{\rm b}=0 }     \nn \\
  =&  \bigg[ \frac{\partial }{ \partial \delta_{\rm b}  }  B_W (k_{W1},k_{W2},k_{W3})  \bigg|_{ \substack{ k_W  \\ \mathrm{fixed} } }  +  \nn \\
      & \quad   \sum_{i=1}^3 \frac{\partial }{ \partial k_{Wi} }  B_W (k_{W1},k_{W2},k_{W3}) \bigg|_{ \substack{ B_W  \\ \mathrm{fixed} } } \frac{\partial k_{Wi}   }{ \partial \delta_{\rm b}} \bigg]_{\delta_{\rm b}=0} \nn \\
 = &   \frac{\partial }{ \partial \delta_{\rm b}  }  B_W (k_{1},k_{2},k_{3})  \bigg|_{\delta_{\rm b}=0}    -  \frac{1}{3}  \sum_{i=1}^3   \frac{\partial  }{ \partial \ln k_{i}  } B( k_{1}, k_{2}, k_{3}).  
\end{align}
The first term in the last line of Eq.~\eqref{eq:dBw_ddeltab_chainrule} encodes the modification of the intrinsic growth due to the long mode.   

In summary, if the global mean is used, the full response function is given by
\begin{align}
\label{eq:response_derivative_full_gb} 
& \frac{\partial  }{\partial \delta_{\rm b} } B (k_1,  k_2,  k_3| \delta_{\rm b} )  \bigg|_{\delta_{\rm b}=0 }  \nn \\
= &    B( k_{1},  k_{2},  k_{3} )  -  \frac{1}{3}  \sum_{i=1}^3   \frac{\partial  }{ \partial \ln k_{i}  } B( k_{1}, k_{2}, k_{3}) \nn \\
& \quad  +   \frac{\partial }{ \partial \delta_{\rm b}  }  B_W (k_{1},k_{2},k_{3})  \bigg|_{\delta_{\rm b}=0} . 
\end{align}
If the local mean is used, with the replacement Eq.~\eqref{eq:global_local_mean_replacement}, there is an additional term  $-3 B $ in Eq.~\eqref{eq:response_derivative_full_gb}.  
These results are similar to the analogous expressions for the power spectrum \cite{Li:2014sga}.  

\section{ The bispectrum response function from theory }
\label{sec:Bk_reponse_SPT_HM}

In this section, we compute the dark matter bispectrum response function using standard perturbation theory (SPT) and then the halo model. The SPT response function is valid in the low $k$ regime, while the halo model will enable us to extend the results to the deeply nonlinear regime.

\subsection{ Coupling of the long and short modes in SPT }
\label{sec:response_SPT_longshortcoupling}

Here we compute the coupling between the long and short modes using SPT (see \cite{PTreview} for a review of SPT).  We see below that the dilation effect and the modification of the growth discussed in Sec.~\ref{sec:Bk_response_effect_general}  appear naturally in SPT.

To obtain the supersample covariance, we need to calculate the linear response function,  i.e.~the first derivative of the bispectrum with respect to the long mode.  Therefore, we only need to compute the modulated density up to first order in $ \delta_{\rm b} $. To evaluate the tree-level bispectrum, second order in the small scale modes is required.

Let us start with the  second order density contrast $\delta^{(2)}$, and we will see shortly that the calculations for  $\delta^{(3)}$ are similar.  In SPT, $\delta^{(2)}$ can be expanded as 
\beq
\label{eq:delta2_SPT}
\delta^{(2)}(\bm{k} ) = \int \frac{ d^3 q }{(2 \pi)^3} F_2(\bm{q}, \bm{k}- \bm{q} ) \delta^{(1)}( \bm{q})  \delta^{(1)}  ( \bm{k}- \bm{q} ), 
\eeq
where $F_2 $ is the coupling kernel
\beq
F_2( \bm{k}_1, \bm{k}_2 ) = \frac{5}{7} + \frac{1}{2} \mu \Big( \frac{k_1}{k_2} +  \frac{k_2}{k_1}  \Big) +  \frac{2}{7} \mu^2, 
\eeq
with $ \mu = \hat{\bm{k}}_1 \cdot \hat{\bm{k}}_2 $.  The convolution integral in Eq.~\eqref{eq:delta2_SPT}, couples $\delta^{(1)}$ of different scales. For example, if both $\delta^{(1)}$ are the small scale  $\delta^{(1)}_{\rm s} $, then it gives the small-scale $\delta_{\rm s}^{(2)} $.  We are particularly interested in the coupling between the long mode  $\delta^{(1)}_{\rm l} $ and the short mode  $\delta^{(1)}_{\rm s} $. Focusing on the long-short coupling, we have
\begin{align}
  \label{eq:delta2_ls}
\delta^{(2)}_{\rm ls}(\bm{k} ) &= 2 \int \frac{ d^3 q }{(2 \pi)^3} F_2(\bm{q}, \bm{k}- \bm{q} ) \delta_{\rm l}^{(1)}( \bm{q})  \delta_{\rm s}^{(1)}  ( \bm{k}- \bm{q} ),
\end{align}
where $\bm{q} $ and $\bm{k}$ represent the long and short modes respectively. As there are poles in $F_2( \bm{q}, \bm{k} - \bm{q}  ) $, we consider the spherically symmetric long wavelength perturbation 
\begin{align}
\label{eq:long_mode_form}
\delta_{\rm l} (\bm{q}) = \frac{ 2\pi^2 \delta_{\rm b} }{ q_{\rm b}^2  }  \Ddel(q -q_{\rm b}) , 
\end{align}
with $q_{b} \ll k$. Eq.~\eqref{eq:long_mode_form} can be obtained by spherically averaging over the angle of $\bm{q}$ in Eq.~\eqref{eq:delta_b_3Dform}, and assuming finite $q_{\rm b} $.   Following \cite{Baldauf:2015vio}, we expand both $ F_2(\bm{q}, \bm{k}- \bm{q} ) $ and $ \delta^{(1)}  ( \bm{k}- \bm{q} ) $ about the long mode  $\bm{q}$. We note that  $ \delta^{(1)} $ is a (Gaussian) random field, so normally it is not differentiable. Crucially, the small scale modes are separated by $k_{\rm F} $ and $q \ll k_{\rm F} $, thus the Taylor expanded value does not interfere the neighboring value and cause a contradiction.

Collecting terms up to order $q^0$, we have
\begin{align}
  \label{eq:delta2_ls}
      \delta^{(2)}_{\rm ls}(\bm{k} ) &  \approx 2 \delta_{\rm b} \int d q \, \Ddel( q - q_{\rm b} ) \int \frac{ d \Omega_q }{4 \pi}      \nn \\
  & \times   \Big\{ \frac{\bm{k} \cdot \bm{q} }{ 2 q^2 }   + \Big[   \frac{3}{14} + \frac{ 2 ( \bm{k}\cdot \bm{q})^2 }{ 7k^2 q^2 } \Big] \Big\}  \nn  \\
      & \times [ \delta_{\rm s}^{(1)} ( \bm{k})  - \bm{q} \cdot \partial_{\bm{k}} \delta_{\rm s}^{(1)} (\bm{k})     ]  \nn  \\
    &=  \Big[ \frac{ 13 }{21 }  \delta^{(1)}(\bm{k}) - \frac{1}{3} \bm{k} \cdot \partial_{\bm{k}} \delta^{(1)}(\bm{k} ) \Big] \delta_{\rm b}   .     
\end {align}
Using Eq.~\eqref{eq:delta2_ls}, up to first order in the long and short modes, we have \cite{Baldauf:2015vio}
\begin{align}
  \label{eq:density2_modulated_SphAve}
\delta_{\rm s}^{(1)}( \bm{k}| \delta_{\rm b}) & = 
\delta^{(1)}_{\rm s} (\bm{k} ) + \delta^{(2)}_{\rm ls}(\bm{k} ) \nn \\
&\approx   \delta^{(1)}_{\rm s} \Big( \bm{k}\big( 1 - \frac{1}{3} \delta_{\rm b} \big)   \Big) +  \frac{ 13 }{21 } \delta_{\rm b} \delta^{(1)}_{\rm s}  (\bm{k}). 
\end{align}
The first term is the dilation effect discussed in Sec.~\ref{sec:Bk_response_effect_general} while the second term is the modification of the small scale growth by the long mode.  This shows that both the modification of the intrinsic growth and the dilation effects are incorporated into SPT automatically, while SPT is normalized with respect to the global mean. 

To first order in $\delta_{\rm b} $ and the short mode, the modulated dark matter power spectrum is given by 
\beq
P(k| \delta_{\rm b}) = P(k) + \delta_{\rm b} \Big[  P(k) - \frac{1}{3} \frac{d P(k)}{d \ln k}   \Big] + \frac{26}{21} \delta_{\rm b} P(k), 
\eeq
where we have split the contributions into the dilation (the term in the square brackets) and the modification of intrinsic growth (last term). We then get the power spectrum response
\beq
\label{eq:Pk_response_SPT_gb}
\frac{ \partial P(k| \delta_{\rm b})    }{ \partial \delta_{\rm b} } \bigg|_{\delta_{\rm b} = 0 } = \frac{47}{21} P(k) -  \frac{1}{3} \frac{d P(k)}{d \ln k}. 
\eeq
If the local mean is used instead, 47/21 is replaced by 5/21 in Eq.~\eqref{eq:Pk_response_SPT_gb}. 


\begin{widetext}
For the tree-level bispectrum response,  we also need the coupling between one long mode and two short modes through the $F_3$  kernel. This long-short-short coupling term is given by
\begin{align}
  \label{eq:delta2_lss}
  \delta_{\rm lss}^{(3)}( \bm{k} )   = &  3 \int \frac{ d^3 q }{ (2 \pi)^3 }  \int \frac{ d^3 k_1 }{ (2 \pi)^3 }  F_3( \bm{q},\bm{k}_1, \bm{k} - \bm{k}_{1} - \bm{q}  ) 
  \delta^{(1)}_{\rm l}(\bm{q} )   \delta^{(1)}_{\rm s}(\bm{k}_1 )   \delta^{(1)}_{\rm s}( \bm{k} - \bm{k}_{1} - \bm{q}  ) \nn  \\
  = & 3 \delta_{\rm b}  \int \frac{ d^3 k_1 }{ (2 \pi)^3 }   \delta^{(1)}_{\rm s}(\bm{k}_1 )    \int d q \Ddel( q - q_{\rm b} ) \int   \frac{ d \Omega_{q} }{ 4 \pi } 
  F_3( \bm{q}, \bm{k}_1, \bm{k} - \bm{k}_1 -  \bm{q})  
   \delta^{(1)}_{\rm s}( \bm{k} - \bm{k}_1 - \bm{q} ) ,
\end{align}
where $\bm{k}$ and $\bm{k}_1$ denote the short modes while  $\bm{q}$ is the long mode.

Analogous to the case for  $\delta_{\rm ls}^{(2)} $, we expand  $ F_3( \bm{q}, \bm{k}_1, \bm{k} - \bm{k}_1 -  \bm{q}) $ to the order $q^0 $ and
\beq
\delta^{(1)}_{\rm s}( \bm{k} - \bm{k}_1 - \bm{q} ) \approx \delta^{(1)}_{\rm s}( \bm{k} - \bm{k}_1 ) - \bm{q} \cdot \partial_{ \bm{k}  }  \delta^{(1)}_{\rm s}( \bm{k} - \bm{k}_1 ).
\eeq
Up to the order $q^0 $, the terms that survive the angular integration are 
\begin{align}
  \label{eq:delta_lss3}
      \delta_{\rm lss}^{(3)}( \bm{k} ) 
  =  - \frac{1}{3} \delta_{\rm b} \int \frac{ d^3 k_1}{(2 \pi)^3 }   F_2( \bm{k}_1,  \bm{k} - \bm{k}_1 )   \delta_{\rm s}^{(1)} ( \bm{k}_1 )  \bm{k} \cdot  \partial_{ \bm{k} }  \delta_{\rm s }^{(1)}( \bm{k} - \bm{k}_1 )   
  +  \delta_{\rm b}  \int \frac{ d^3 k_1}{(2 \pi)^3 }  A_0(\bm{k}_1, \bm{k} - \bm{k}_1 )  \delta_{\rm s}^{(1)}(\bm{k}_1 ) \delta_{\rm s}^{(1)}( \bm{k} - \bm{k}_1 ), 
\end{align} 
where $A_0$ denotes
\begin{align}
  \label{eq:A0_term}
A_0( \bm{k}_1, \bm{k} - \bm{k}_1 )& = \frac{89}{126} + \frac{11}{ 21 }\bm{k}_1 \cdot ( \bm{k} - \bm{k}_1 ) \Big( \frac{1}{ k_1^2 } + \frac{ 1}{ | \bm{k} - \bm{k}_1|^2 }  \Big) +  \frac{ 23 }{63 } \frac{ [ \bm{k}_1 \cdot ( \bm{k} - \bm{k}_1 ) ]^2 }{k_1^2 | \bm{k} - \bm{k}_1  |^2  } \nn \\
& -\frac{1}{12} \Big[ \frac{ k_1^2 }{| \bm{k} - \bm{k}_1 |^2}  + \frac{| \bm{k} - \bm{k}_1 |^2}{ k_1^2 }    \Big]  + \frac{ [ \bm{k}_1 \cdot (  \bm{k} - \bm{k}_1  )]^2  }{ 6  } \Big[  \frac{1}{ k_1^4 }+ \frac{1}{|\bm{k} - \bm{k}_1 |^4 }  \Big]   \nn \\
& +   \frac{2}{21}   \frac{  [ \bm{k}_1 \cdot (  \bm{k} - \bm{k}_1  )]^3    }{ k_1^2 |\bm{k} - \bm{k}_1 |^2    } \Big( \frac{1}{ k_1^2 } + \frac{1}{|\bm{k} - \bm{k}_1 |^2}     \Big).    
\end{align}
We have symmetrized the kernel $A_0$.  The first term in Eq.~\eqref{eq:delta_lss3} is due to the product of the $q^{-1}$-order term in $ F_3( \bm{q}, \bm{k}_1, \bm{k} - \bm{k}_1 -  \bm{q}) $  with  the gradient term  $-\bm{q} \cdot \partial_{\bm{k}}  \delta^{(1)}_{\rm s}( \bm{k} - \bm{k}_1 )$; and the second term results from  the product of the $q^0$-order term in $ F_3( \bm{q}, \bm{k}_1, \bm{k} - \bm{k}_1 -  \bm{q}) $ with  $ \delta^{(1)}_{\rm s}( \bm{k} - \bm{k}_1 )$.

By simply adding and subtracting the term
\beq
- \frac{1}{3} \delta_{\rm b} \int \frac{ d^3 k_1}{(2 \pi)^3 }  [  \bm{k} \cdot \partial_{\bm{k}}   F_2( \bm{k}_1,  \bm{k} - \bm{k}_1 ) ]  \delta_{\rm s}^{(1)}( \bm{k}_1 )   \delta_{\rm s}^{(1)} ( \bm{k} - \bm{k}_1 ),
\eeq
we can write  Eq.~\eqref{eq:delta_lss3} as
\begin{align}
  \delta_{\rm lss}^{(3)}( \bm{k} )  =  - \frac{1 }{3} \delta_{\rm b} \bm{k} \cdot  \partial_{\bm{k}} \delta_{\rm s}^{(2)} (\bm{k} ) +
   \delta_{\rm b}  \int \frac{ d^3 k_1}{(2 \pi)^3 }  A(\bm{k}_1, \bm{k} - \bm{k}_1 )   \delta_{\rm s}^{(1)}(\bm{k}_1 ) \delta_{\rm s}^{(1)}( \bm{k} - \bm{k}_1 ),
\end{align}
where $A$ reads
\beq
\label{eq:Akernel}
A ( \bm{k}_1, \bm{k} - \bm{k}_1 ) = \frac{ 151}{ 126 } F_2(\bm{k}_1, \bm{k} - \bm{k}_1)   + \frac{5}{ 126 } G_2(\bm{k}_1, \bm{k} - \bm{k}_1) ,   
\eeq
and $ \delta_{\rm s}^{(2)} $ is defined similar to Eq.~\eqref{eq:delta2_SPT} except with $\delta^{(1)}$ replaced by  $\delta^{(1)}_{\rm s} $.   $G_2$ is the velocity divergence  kernel
\beq
G_2( \bm{k}_1, \bm{k}_2 ) = \frac{3}{7} + \frac{1}{2} \mu \Big( \frac{k_1}{k_2} +  \frac{k_2}{k_1}  \Big) +  \frac{4}{7} \mu^2.  
\eeq
Note that the three types of terms in Eq.~\eqref{eq:A0_term} are canceled out in Eq.~\eqref{eq:Akernel}, and $A$ can be solely written in terms of $F_2$ and $G_2$.

Therefore up to first order in the long mode and second order in the short modes, the small scale mode reads
\begin{align}
\label{eq:delta2s_long}
\delta_{\rm s}^{(2)}(\bm{k} | \delta_{\rm b} ) &= 
\delta_{\rm s}^{(1)}( \bm{k} ) +  \delta_{\rm s}^{(2)}( \bm{k} ) - \frac{1}{3 }\delta_{\rm b} \bm{k} \cdot \partial_{\bm{k} } \delta_{\rm s}^{(1)} ( \bm{k}  ) - \frac{1}{3 }\delta_{\rm b} \bm{k} \cdot \partial_{\bm{k} }\delta_{\rm s}^{(2)} ( \bm{k} )  + \frac{13}{21} \delta_{\rm b} \delta_{\rm s}^{(1)}(\bm{k})  \nn \\
& \qquad   + \delta_{\rm b} \int \frac{ d^3k_1}{(2 \pi)^3  } A(\bm{k}_1,\bm{k} - \bm{k}_1)    \delta_{\rm s}^{(1)}(\bm{k}_1)    \delta_{\rm s}^{(1)}( \bm{k} - \bm{k}_1) \nn \\
&\approx  \delta_{\rm s}^{(1)}\Big( \bm{k}\big(1 - \frac{1}{3} \delta_{\rm b}\big) \Big) + \delta_{\rm s}^{(2)}\Big( \bm{k}\big(1 - \frac{1}{3} \delta_{\rm b}\big) \Big)  +  \delta_{\rm b}     \Big[\frac{13}{21} \delta_{\rm s}^{(1)}( \bm{k}) +  \int \frac{ d^3k_1}{(2 \pi)^3  } A(\bm{k}_1,\bm{k} - \bm{k}_1)    \delta_{\rm s}^{(1)}(\bm{k}_1)    \delta_{\rm s}^{(1)}( \bm{k} - \bm{k}_1)      \Big] .
\end{align}
In the last line, the first two terms are the dilation terms, while the last term is the modification of the intrinsic growth up to second order in the short mode. If the local mean is used, there is an additional overall factor $1/(1 + \delta_{\rm b}) $ in Eq.~\eqref{eq:delta2s_long}.

We are now in a position to compute the modulated bispectrum. The dilation part of the bispectrum can be obtained as
\begin{align}
   &  \Big\langle  \delta_{\rm s}^{(1)}\Big( \bm{k}_1\big(1 - \frac{1}{3} \delta_{\rm b}\big) \Big)   \delta_{\rm s}^{(1)}\Big( \bm{k}_2\big(1 - \frac{1}{3} \delta_{\rm b}\big) \Big)  \delta_{\rm s}^{(2)}\Big( \bm{k}_3\big(1 - \frac{1}{3} \delta_{\rm b}\big) \Big)   \Big\rangle  +  \,2 \, \mathrm{cyc.}   \nn \\
= & 2 F_2\Big( \bm{k}_1( 1 - \frac{1}{3} \delta_{\rm b}),   \bm{k}_2( 1 - \frac{1}{3} \delta_{\rm b})        \Big) P \Big( k_1( 1 - \frac{1}{3} \delta_{\rm b}) \Big) P \Big( k_2( 1 - \frac{1}{3} \delta_{\rm b}) \Big)  \,  (2 \pi)^3 \Ddel\Big( (1 - \frac{1 }{ 3 }\delta_{\rm b}) ( \bm{k}_1 + \bm{k}_2 +  \bm{k}_3 )\Big)     +  \,2 \, \mathrm{cyc.}   \nn \\ 
\approx & (1 + \delta_{\rm b}) B_{\rm m} \Big( k_1( 1 - \frac{1}{3} \delta_{\rm b}), k_2( 1 - \frac{1}{3} \delta_{\rm b}), k_3( 1 - \frac{1}{3} \delta_{\rm b}) \Big) \,   (2 \pi)^3\Ddel( \bm{k}_1 + \bm{k}_2 +  \bm{k}_3 ) \nn \\
\approx & \Big[ (1 + \delta_{\rm b } ) B_{\rm m}(k_1,k_2,k_3)  - \frac{1}{3} \delta_{\rm b} \sum_{j=1}^3 \frac{ d  }{ d \ln k_{j } } B_{\rm m } ( k_1,k_2,k_3 )\Big]  (2 \pi)^3\Ddel( \bm{k}_1 + \bm{k}_2 +  \bm{k}_3 )   ,  
\end{align}
where $ B_{\rm m }$ is the tree-level dark matter bispectrum
\beq
B_{\rm m}( k_1,k_2,k_3 ) = 2 F_2( \bm{k}_1, \bm{k}_2 ) P(k_1) P(k_2) +  \,2 \, \mathrm{cyc.} 
\eeq
Including the growth enhancement part, the tree-level bispectrum up to first order in the long mode reads
\begin{align}
  \label{eq:B_SPT_modulated}
  B(k_1,k_2,k_3| \delta_{\rm b} ) = &  \Big(1 + \frac{433}{ 126 } \delta_{\rm b } \Big) B_{\rm m}(k_1,k_2,k_3)  + \frac{5}{126}  \delta_{\rm b} B_{G_2} (k_1,k_2,k_3)         - \frac{1}{3} \delta_{\rm b} \sum_{j=1}^3 \frac{ d  }{ d \ln k_{j } } B_{\rm m } ( k_1,k_2,k_3 ) , 
\end{align}
where  $B_{G_2}$ denotes
\beq
 B_{G_2} (k_1,k_2,k_3) = 2 G_2(k_1,k_2)P(k_1)P(k_2)  +  \,2 \, \mathrm{cyc.} 
\eeq
The bispectrum response function is then given by
\begin{align}
  \label{eq:Bk_response_SPT_global}
   \frac{\partial }{ \partial \delta_{\rm b} }  B(k_1,k_2,k_3| \delta_{\rm b} ) \bigg|_{\delta_{\rm b}= 0 }  = \frac{433}{126} B_{\rm m}(k_1,k_2,k_3) + \frac{5}{126}  B_{G_2}(k_1,k_2,k_3) -  \frac{1}{3}  \sum_{j=1}^3 \frac{ d  }{ d \ln k_{j } } B_{\rm m } ( k_1,k_2,k_3 ) . 
\end{align}

If the local mean is used, there is an extra term  $-3 \delta_{\rm b}  B_{\rm m} $ in Eq.~\eqref{eq:B_SPT_modulated} and hence the factor $443/126$ in  Eq.~\eqref{eq:Bk_response_SPT_global} is replaced by $55/126$.


\end{widetext}

\begin{figure}[!htb]
\begin{center}
\includegraphics[width=\linewidth]{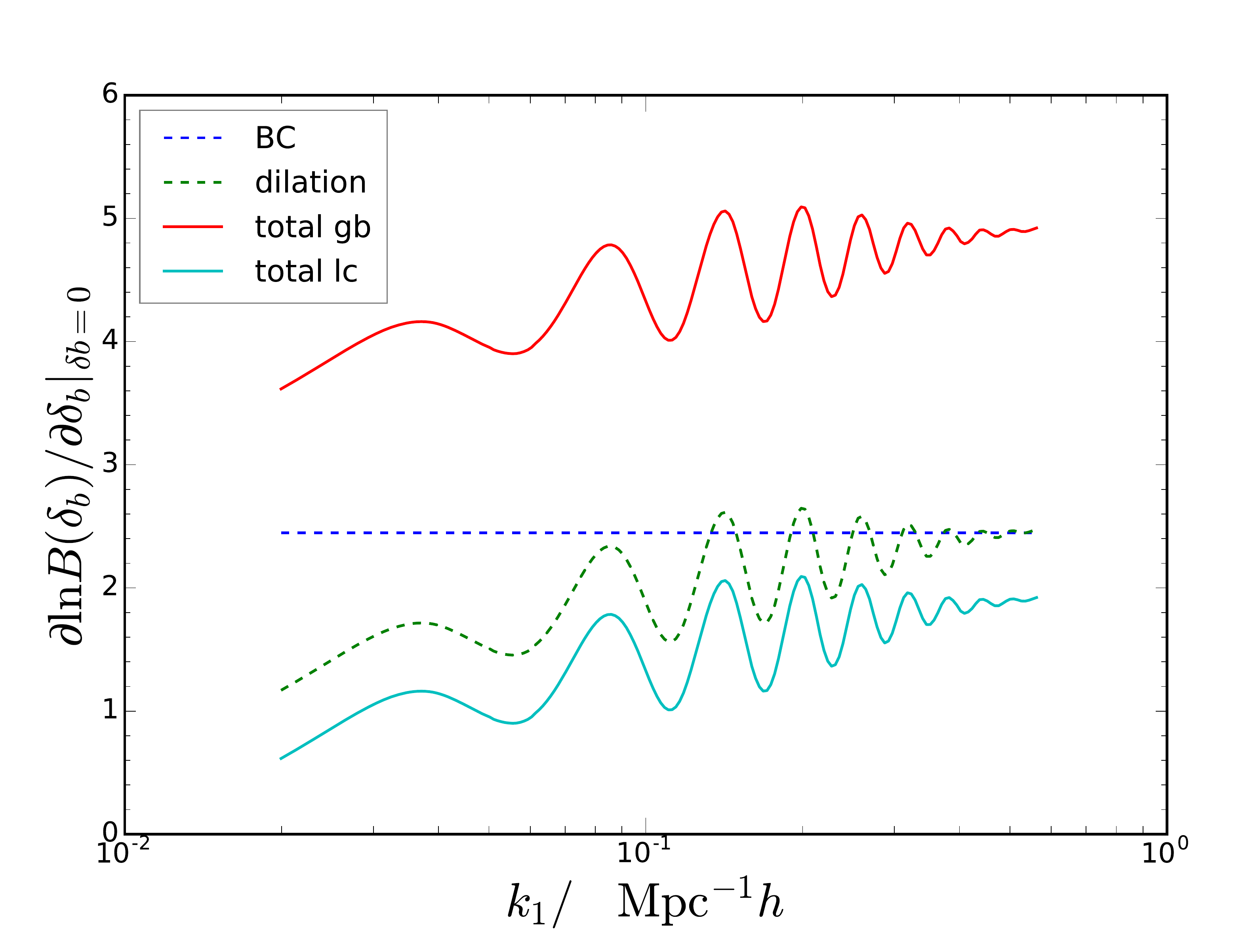}
\caption{ The components of the dark matter bispectrum response function normalized by the bispectrum for the equilateral triangle configuration. The beat coupling (dashed, blue) and the dilation effect (dashed, green) are  plotted separately. The total response obtained using local mean (solid, cyan) is significantly reduced relative to the global mean case (solid, red).}
\label{fig:Bresponse_component}
\end{center}
\end{figure}

In Fig.~\ref{fig:Bresponse_component}, we plot the components of the dark matter bispectrum response function for the equilateral triangle configuration. Both the dilation effect and the modification of the growth add up in the response function. We also compare the cases of a global mean and a local mean. For galaxy surveys, a local mean is used, while for weak lensing a global mean is used.  The response function is significantly reduced for the case of a local mean.

\vspace{-0.8cm}

Here spherical averaging over the angle of the long mode is used, so the effects of the large-scale tidal perturbations are averaged out.  In Appendix \ref{sec:SSC_tides}, we generalize the computations to include the long-wavelength tidal contributions on the small-scale matter bispectrum using SPT; in this way, we are able to derive the bispectrum response function to the tides [Eq.~\eqref{eq:response_tide}]. A key difference of the tidal response function from the density one is that it is anisotropic. This will be important for the final tidal contribution to the supersample covariance.

\subsection{ Bispectrum response function from halo model }
\label{sec:bisp_response_HM}

The response function can also be computed using the halo model approach \cite{CooraySheth,Peacock:2000qk,Seljak:2000gq,Scoccimarroetal2001}. In the halo model, all the dark matter is assumed to reside in halos of different masses. The halo model provides a reasonably accurate phenomenological method to extend the polyspectrum to high $k$.

It is instructive to first review the computation of the halo model power spectrum response function \cite{Takada:2013bfn}. The halo model dark matter power spectrum reads
\beq
\label{eq:Pk_HM}
P_{\rm HM}(k) = [ I^1_1(k) ]^2 P(k) + I^0_2(k), 
\eeq
where the first term is the 2-halo term, which describes the correlation of dark matter in two different halos, and the second term is the 1-halo term, which describes the correlation in the same halo.  Following \cite{Cooray:2000ry} we use the general notation $I_\mu^\beta$
\begin{align}
I_\mu^\beta(k_1,k_2,...,k_\mu) &\equiv \int d  M      \bigg[     \left( \frac{M} {\bar \rho_m} \right)^\mu b_\beta(M) n(M)  \nn \\
 &\times   u_M(k_1)  u_M(k_2) \dots u_M(k_\mu) \bigg],
\end{align}
where $M$ is the halo mass, $n$ is the halo mass function,  $b_\beta$ is the peak-background split bias of order $\beta$, and $ u_M(k_\mu)$ is the dimensionless Fourier transform of the halo density profile normalized such that $ u_M(0)=1$. We use the NFW halo profile \cite{Navarro:1996gj} with the concentration relation given in \cite{CooraySheth}. The Sheth-Tormen mass function \cite{ShethTormen} and the peak-background split bias derived from it \cite{Scoccimarroetal2001} are adopted. We assume only linear bias and hence $b_\beta = 0$ for $\beta\geq 2$.

In the standard halo model formula Eq.~\eqref{eq:Pk_HM}, the long mode vanishes. To compute the response function, we imagine the long mode is turned on and it modulates $P$, and also the mass function and the bias, while we assume that the halo profile is not affected by the long mode.  The response of the mass function and the bias to the long mode can be derived using the relation
\begin{align}
  \label{eq:nb_response_PBS}
\frac{\partial I_\mu^\beta}{\partial \delta_{\rm b}} \bigg|_{\delta_{\rm b}=0} & = \int d M \bigg[ \left(\frac{M}{\bar \rho_m}\right)^\mu  \frac{\partial  }{\partial \delta_{\rm b}}  [ b_\beta(M) n( M) ] \nonumber \\
&\times  u_M(k_1) u_M(k_2) ... u_M(k_\mu) \bigg]_{\delta_{\rm b} = 0} = I_\mu^{\beta+1} ,
\end{align}
where we have used the fact that the peak-background split bias  $b_\beta$ is the response of mass function to $\delta_{\rm b}$ at order $\beta$ \cite{Mo:1995cs,Schmidt:2012ys} 
\beq
b_\beta(M) = \left.\frac{1}{n(M)}\frac{\partial^\beta n(M)}{\partial \delta_{\rm b}^\beta}\right |_{\delta_{\rm b} = 0}.
\eeq

With this setup, it is easy to see that the power spectrum response function is given by \cite{Takada:2013bfn}
\beq
\label{eq:PHM_response}
\frac{\partial P_{\rm HM}(k | \delta_{\rm b} ) }{\partial \delta_{\rm b} } \bigg|_{\delta_{\rm b}= 0}  \approx [ I_1^1(k) ]^2 \frac{\partial P( k | \delta_{\rm b} )  }{\partial \delta_{\rm b} }\bigg|_{\delta_{\rm b}= 0}  + I^1_2(k), 
\eeq
where we have dropped a term proportional to $I^2_1 $ because it is small compared to the ones we keep.  The perturbative  power spectrum response function is given by Eq.~\eqref{eq:Pk_response_SPT_gb}.  For the case of local mean, there is an additional term $-2 P_{\rm HM} $ in Eq.~\eqref{eq:PHM_response}.

\begin{figure}[!tb]
\begin{center}
\includegraphics[width=\linewidth]{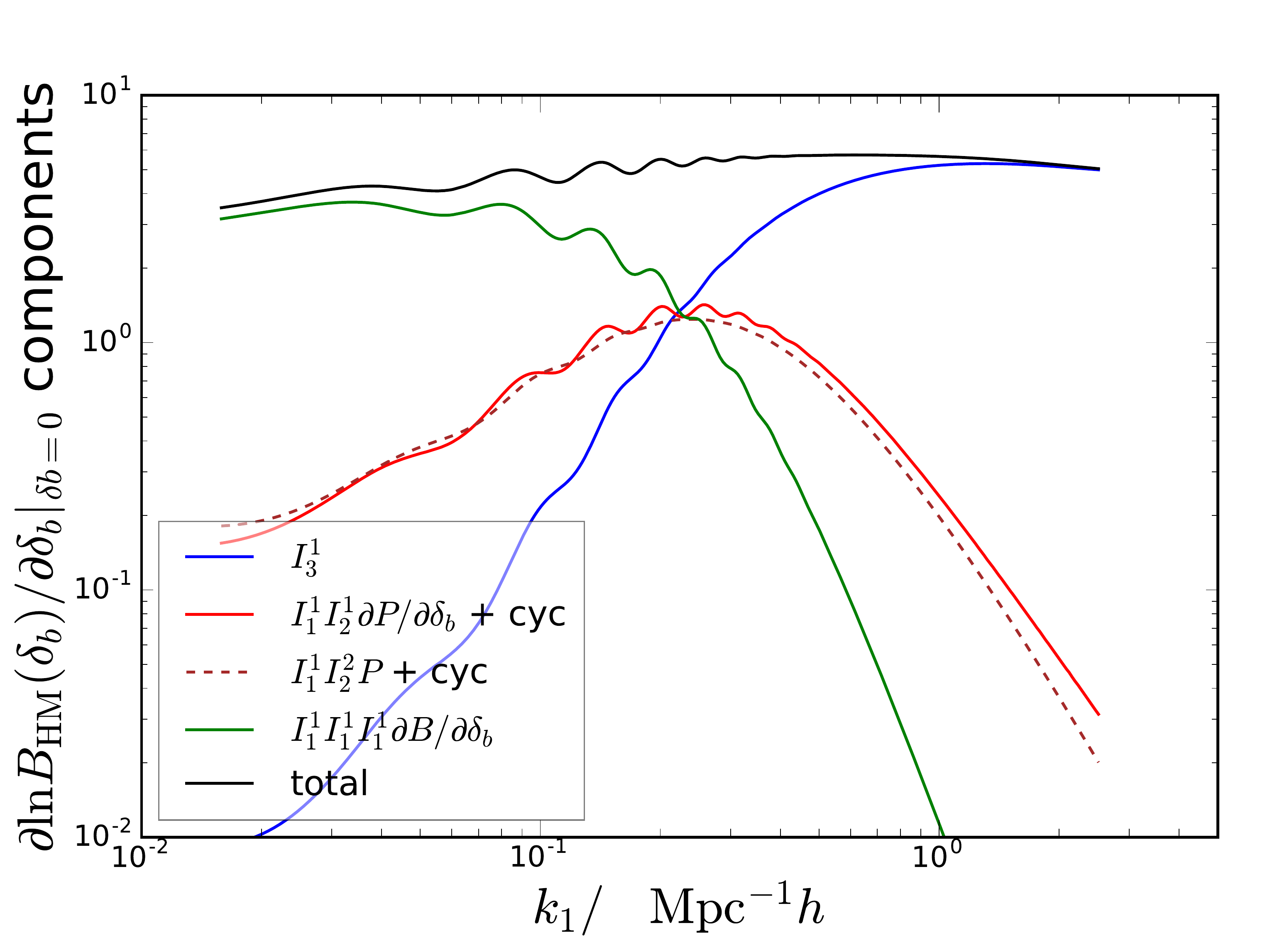}
\caption{  The bispectrum response function (solid, black)  computed using the halo model. The individual terms are also shown: the term due to the 3-halo term (red, solid), that due to the 1-halo term (blue, solid), and the two dominant terms from the 2-halo term [the perturbative power spectrum response term (red, solid) and the term involving $b_2$ (brown, dashed)].  Results for the equilateral triangle configurations at $z=0 $ are shown.}
\label{fig:Bkresponse_HaloModel_111_simple_BkNorm}
\end{center}
\end{figure}

For the case of the power spectrum, Ref.~\cite{Takada:2013bfn} explicitly checked that the supersample covariance computed using the halo model trispectrum agrees with the supersample covariance formula results [the analog of Eq.~\eqref{eq:CB_SSC}] calculated using the halo model power spectrum response Eq.~\eqref{eq:PHM_response}.  As it is formidable to check the results using the halo model 6-point function (for a glimpse of its complexity,  see the full 6-point function in the Poisson model in \cite{Chan:2016ehg}), here we directly compute the bispectrum response function using the halo model.

In the language of the halo model,  the dark matter bispectrum reads \cite{Scoccimarroetal2001} 
\begin{align}
  \label{eq:BHM}
B_{\rm HM}(k_1,k_2,k_3)	& = B_{\rm 1h}(k_1,k_2,k_3) + B_{\rm 2h}(k_1,k_2,k_3) \nonumber  \\
			& + B_{\rm 3h}(k_1,k_2,k_3) , 
\end{align}
where
\begin{align}
B_{\rm 1h}(k_1,k_2,k_3) &=  I_3^0(k_1,k_2,k_3),   \\
B_{\rm 2h}(k_1,k_2,k_3) &=  I_1^1(k_1)I_2^1(k_2,k_3)P(k_1) + 2 \, {\rm cyc.},  \\
B_{\rm 3h}(k_1,k_2,k_3) &=  I_1^1(k_1) I_1^1(k_2) I_1^1(k_3) B_{\rm PT}(k_1,k_2,k_3),
\end{align}
are the 1-, 2-, and 3-halo terms, and $ B_{\rm PT}$ denotes the bispectrum from the perturbation theory. The 1-, 2-, and 3-halo terms describe the situations in which  all three points are in the same halo, only two of the points are in the same halo, and none of them are in the same halo, respectively. We follow the same prescription as for the case of the power spectrum.  In particular, we also first assume only a linear bias and  $b_\beta = 0$ for $\beta\geq 2$. In this case, $B_{\rm PT}$ is simply $B_{\rm m}$.

With the assumption that the presence of long mode modulates $P$ and $B_{\rm PT} $, and also the mass function and the bias while the halo profile is not affected, we can write down the halo model bispectrum response function 
\begin{align}
  \label{eq:BHM_response}
  & \frac{\partial}{\partial \delta_{\rm b}}  B_{\rm HM}(k_1,k_2,k_3| \delta_{\rm b})  \bigg|_{\delta_{\rm b} = 0 }  \nn \\
  \approx  &  I_1^1(k_1)  I_1^1(k_2)  I_1^1(k_3) \frac{\partial B_{\rm PT}(k_1,k_2,k_3| \delta_{\rm b}) }{\partial \delta_{\rm b} }\bigg|_{\delta_{\rm b} = 0 }  \nn \\
  & + \bigg[ I^1_1(k_1) I^2_2(k_2,k_3) P(k_1)    \nn \\
  &  + I_1^1(k_1) I_2^1(k_2,k_3) \frac{\partial P(k_1| \delta_{\rm b} )}{\partial \delta_{\rm b} }\bigg|_{\delta_{\rm b} = 0 }  \bigg] + 2 \, \mathrm{cyc}. \nn \\
& +  I_3^1(k_1,k_2,k_3) . 
\end{align}
The response function  $[ \partial B_{\rm PT}(\delta_{\rm b}) / \partial \delta_{\rm b}  ]_{\delta_{\rm b} = 0 } $ is given by Eq.~\eqref{eq:Bk_response_SPT_global}, and the perturbative power spectrum response function is given by Eq.~\eqref{eq:Pk_response_SPT_gb}.  If the local mean is used instead, there is an additional term $-3 B_{\rm HM}$ in Eq.~\eqref{eq:BHM_response}.

In Fig.~\ref{fig:Bkresponse_HaloModel_111_simple_BkNorm},   we plot the bispectrum response function at $z=0 $ for the case of the  global mean.  At large scales, for $k \lesssim 0.4 \hOMpc $, the 3-halo contibution dominates, while at small scales, for $k\gtrsim 0.6 \hOMpc $, the 1-halo contribution  $I_3^1$, becomes dominant. In the scales shown, the 3-halo contribution is essentially the same as the bispectrum response function computed using perturbation theory.  However, as we can see in Fig. \ref{fig:Bkresponse_HaloModel_111_simple_BkNorm}, on large scale, the halo model prediction still differs from the perturbation theory results, e.g.~at $k \sim 0.02 \hOMpc $, the halo model result exceeds that of perturbation theory by 10\%. At large scales, it is well known that the standard halo model formalism predicts unphysical shot noise between matter and halo (see \cite{Ginzburg:2017mgf} for a recent attempt to resolve this issue). Our result indicates that there seems to be another artifact of the halo model at large scales. However, this effect is negligible when we compare the covariance predictions with the numerical results later on.

The terms in the square brackets in Eq.~\eqref{eq:BHM_response} are small in the low and high $k$ regimes, but they are not negligible at the transition scales. Besides the power spectrum response function term, we have also kept a term involving $b_2$.  From Eq.~\eqref{eq:nb_response_PBS},  we see that although we have limited ourselves to $b_1$ only, higher order bias terms are generated by the response derivative. As we see in Fig.~\ref{fig:Bkresponse_HaloModel_111_simple_BkNorm}, this term is comparable to the perturbative power spectrum response function term, and thus we keep it as well.  We have checked that all of the other  terms generated by the response derivative  are negligible except this one.

As we have included one of the $b_2$-terms generated, we need to check our starting assumption by including only the $b_1$-terms. In the 3-halo term, there is another possible term $I^1_1(k_1) I^1_1(k_2) I^2_1(k_3) P(k_1)P(k_2)  + 2 \, \mathrm{cyc.}$ We can estimate the importance of this term by differentiating this term directly with respect to the long mode. This is not strictly correct as this term is obtained by setting the long mode to zero; however, for the purpose of estimation, it is sufficient. We find that this term is indeed small compared to the ones that we included.

\begin{figure*}[!htb]
\begin{center}
\includegraphics[width=\linewidth]{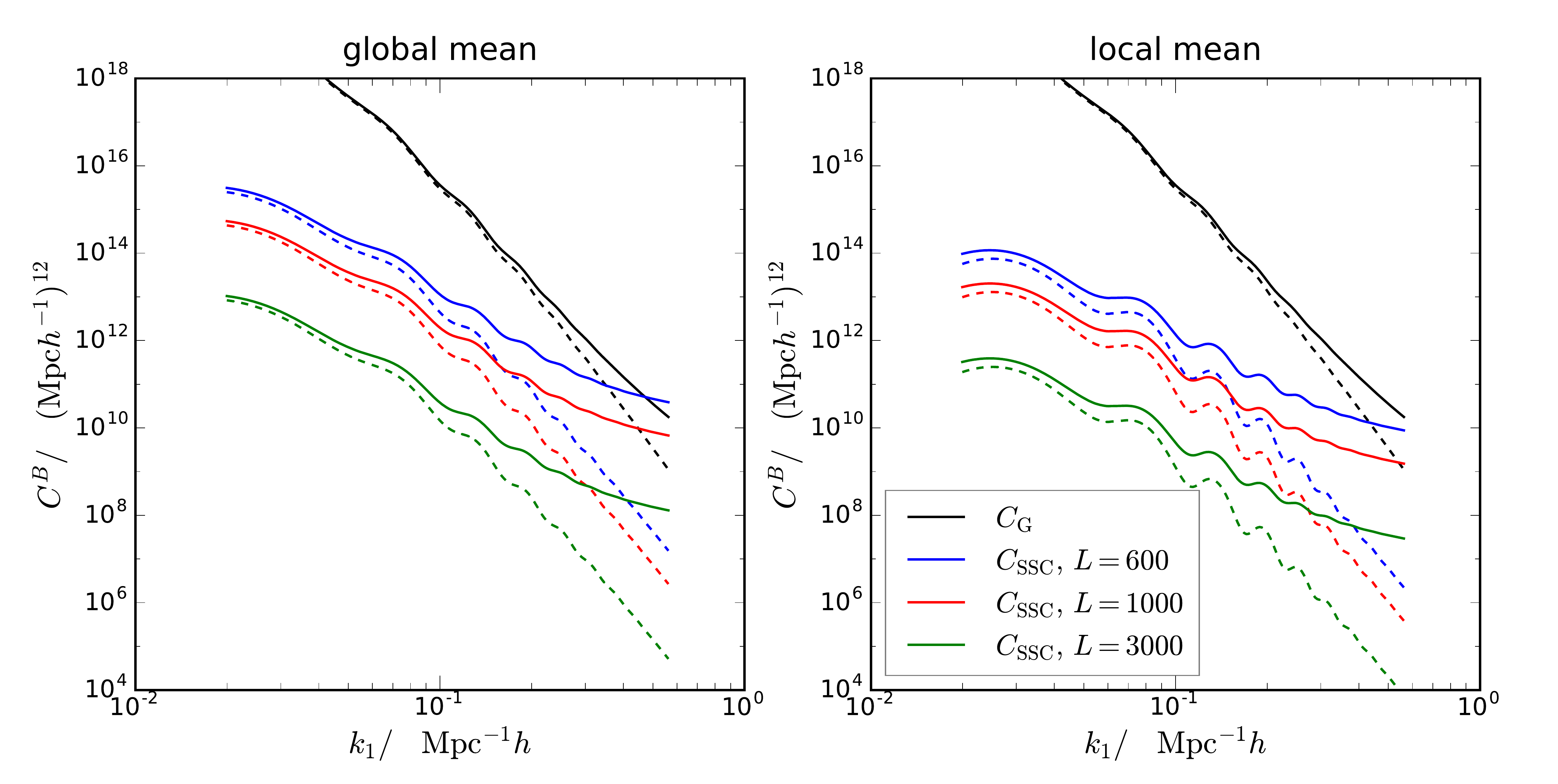}
\caption{ The diagonal elements of the  supersample bispectrum covariances for a suite of values of box size $L$  are  compared with the Gaussian covariance (black, dashed line for using the linear power spectrum, and black, solid line for using the halo model power spectrum). The cases for the global mean (left) and local mean (right) are shown. The colorful solid lines show the halo model prediction and the dashed lines (same color) show the perturbation theory prediction. Equilateral triangle configurations at $z=0$ are shown.  }
\label{fig:CB_SSC_response_111}
\end{center}
\end{figure*}

\begin{figure}[!tb]
\begin{center}
\includegraphics[width=\linewidth]{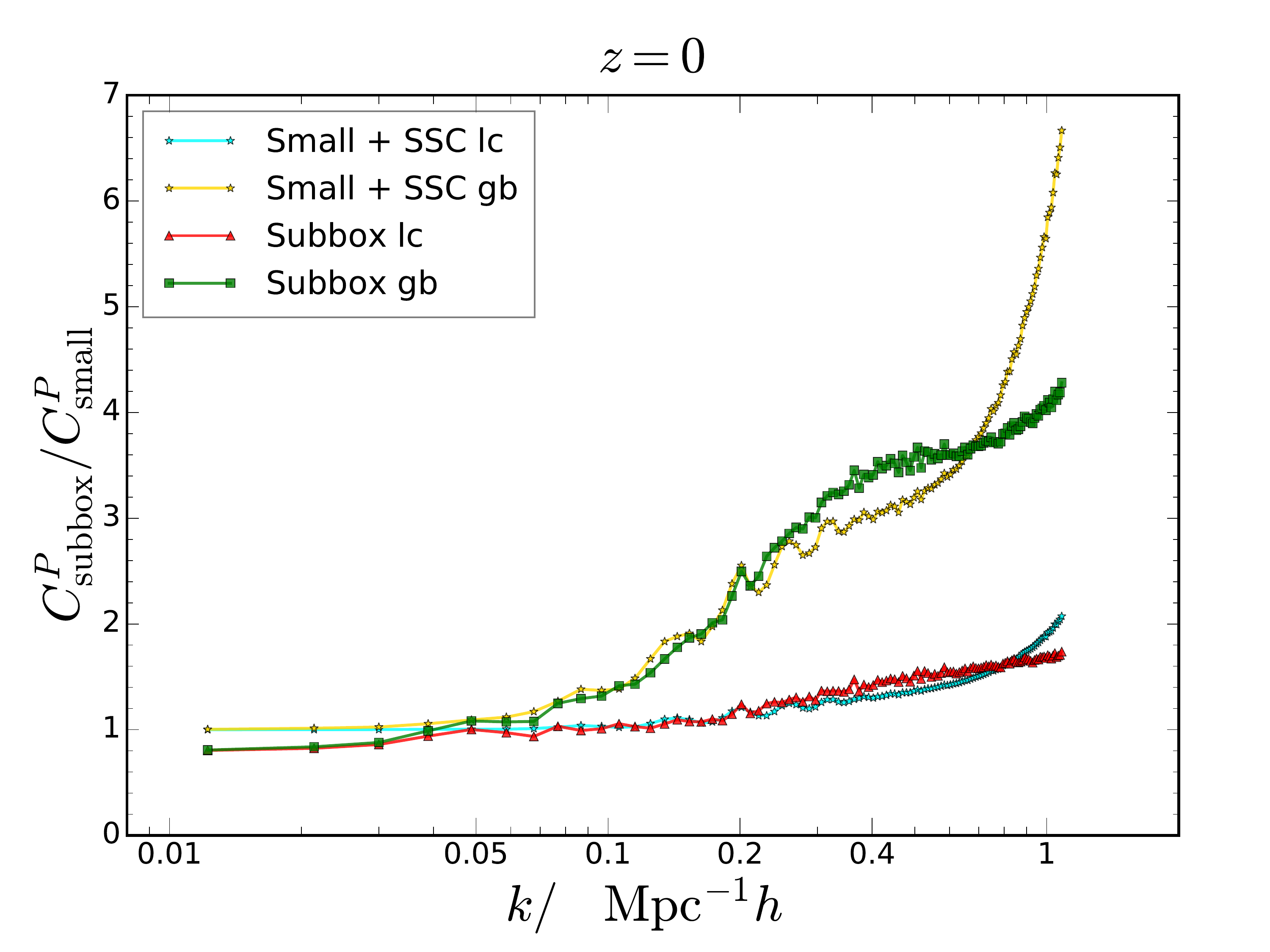}
\caption{ The ratio between the diagonal elements of the power spectrum covariance measured from the subbox and periodic box setup. The results for the subbox setup with the global mean (green, squares) and local mean (red, triangles) are compared. The predictions using the halo model response function are also shown (yellow stars for global mean and cyan stars for local mean).  }
\label{fig:PkCovDiag_ratio_subbox_small_SSC_z0}
\end{center}
\end{figure}


\begin{figure*}[!htb]
\begin{center}
\includegraphics[width=0.45\textwidth]{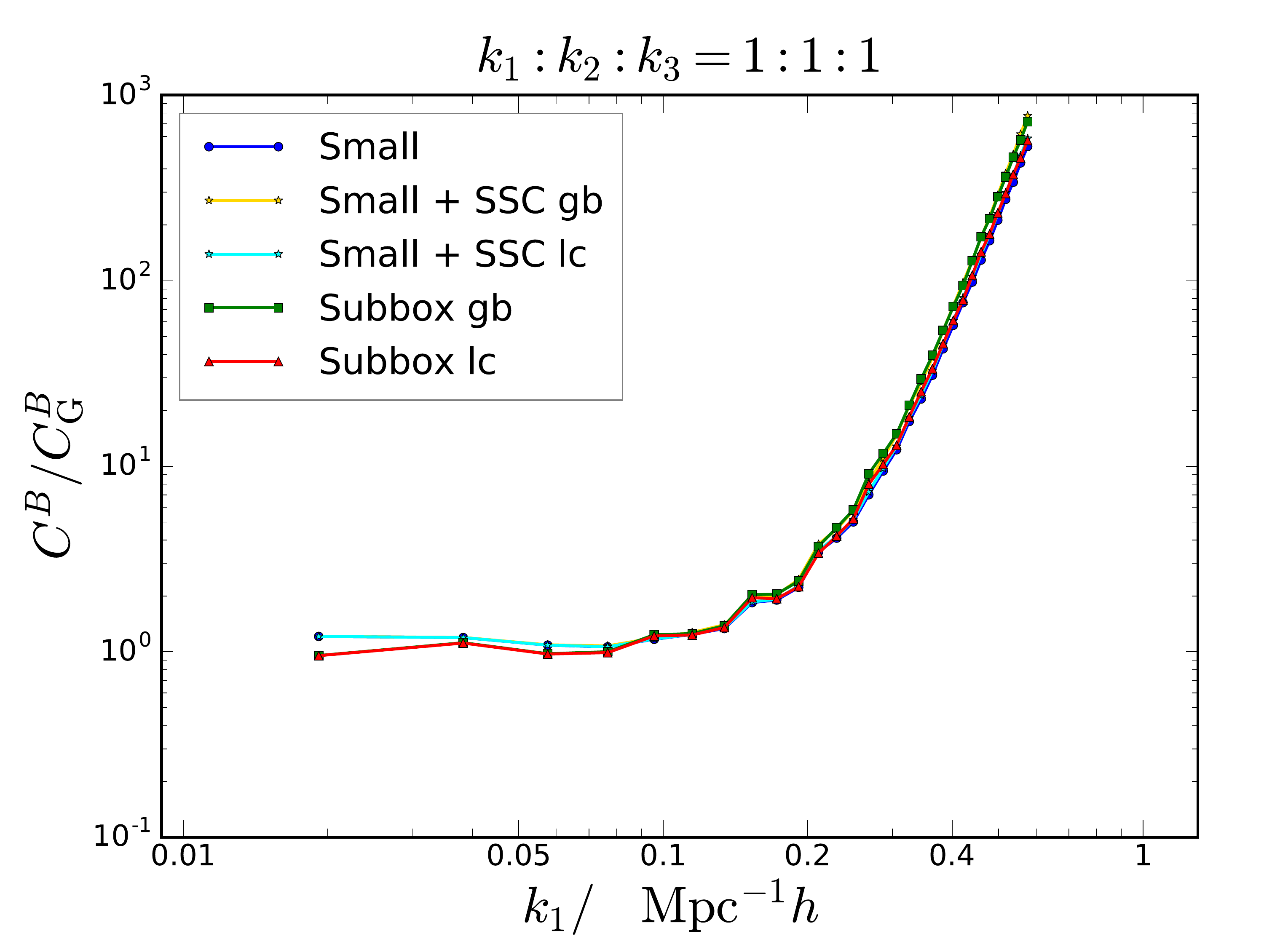}
\includegraphics[width=0.45\textwidth]{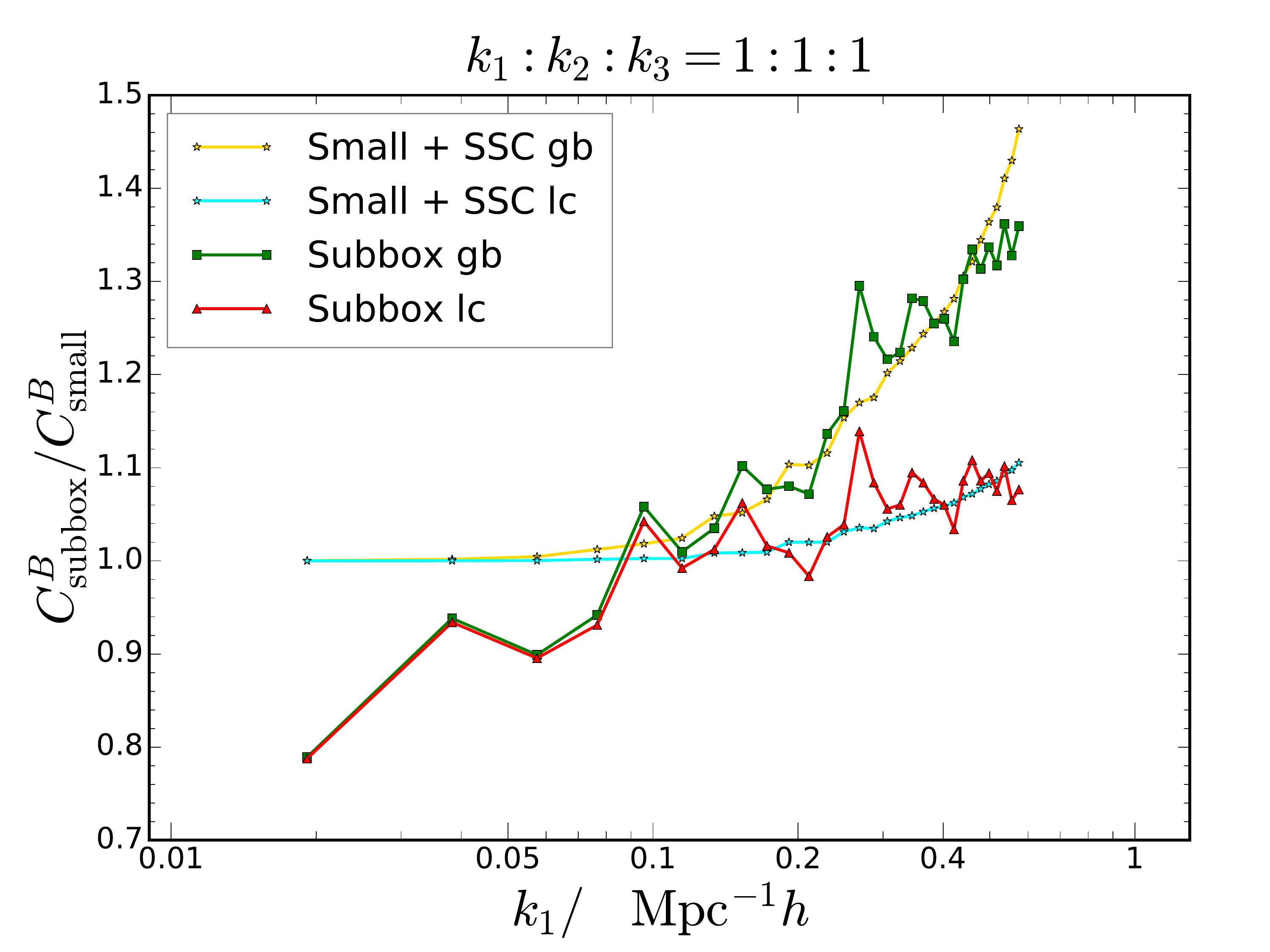}
\caption{ The left panel shows the diagonal element of the dark matter bispectrum covariance for the equilateral triangle configurations at $z=0$, normalized by the Gaussian covariance.  The covariances  measured from the periodic box (blue, circles), and the subbox setup with the global (green, squares) and local mean (red, triangles) are compared.  The supersample covariance predictions for the global (yellow, stars) and local mean (cyan, stars) are shown.   In the right panel, the ratios between various covariances and the one measured from the small set of simulations are plotted. }
\label{fig:diag_cov_mat_small_subbox_111_SSC}
\end{center}
\end{figure*}

\begin{figure}[!htb]
\begin{center}
\includegraphics[width=0.9\linewidth]{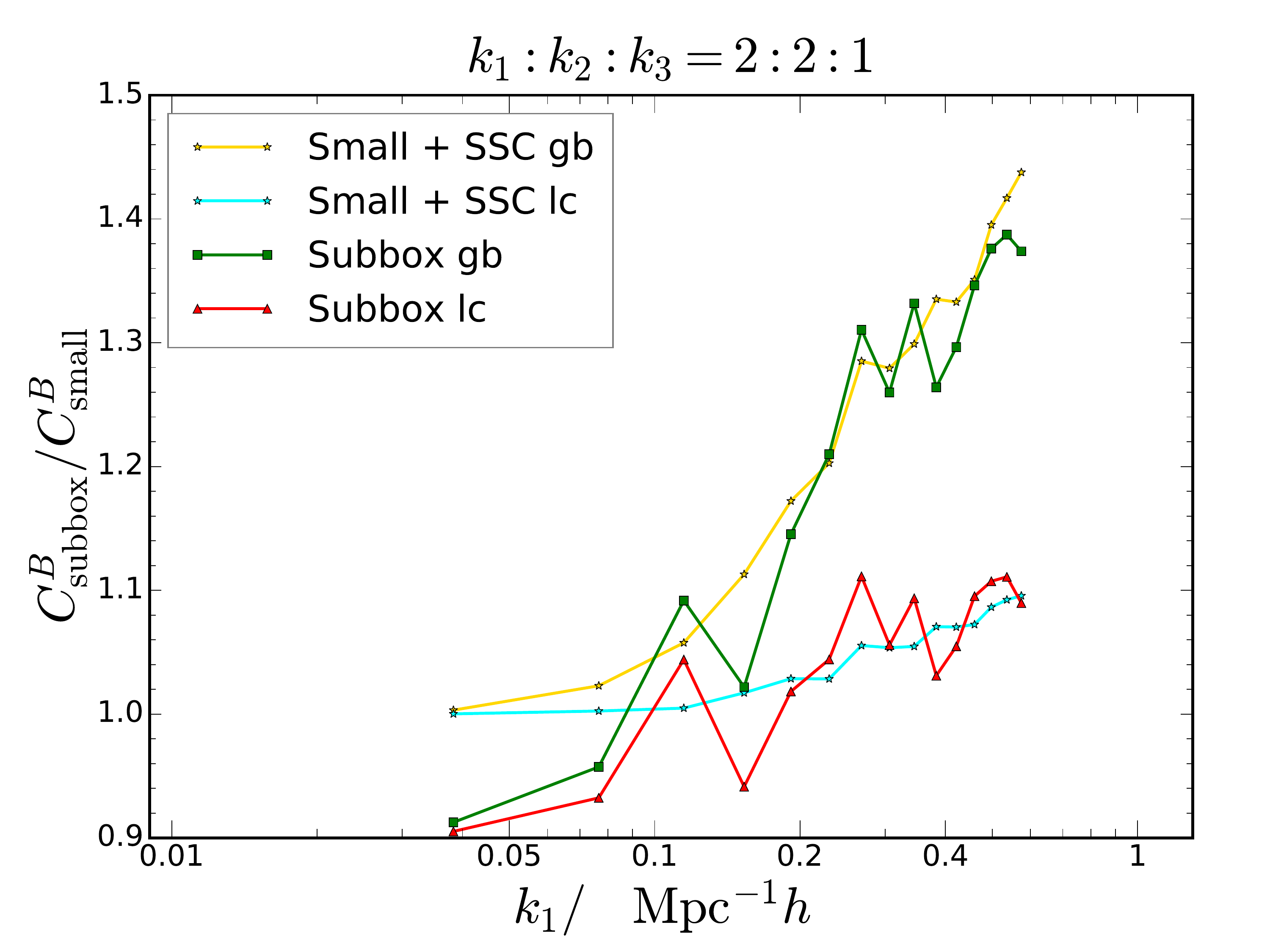}
\caption{ Similar to the right panel of Fig.~\ref{fig:diag_cov_mat_small_subbox_111_SSC}, except for the isosceles triangle of the shape $k_1: k_2 :k_3 = 2:2:1 $. } 
\label{fig:diag_cov_mat_small_subbox_221_SSC}
\end{center}
\end{figure}

\begin{figure*}[!htb]
\begin{center}
\includegraphics[width=0.9\linewidth]{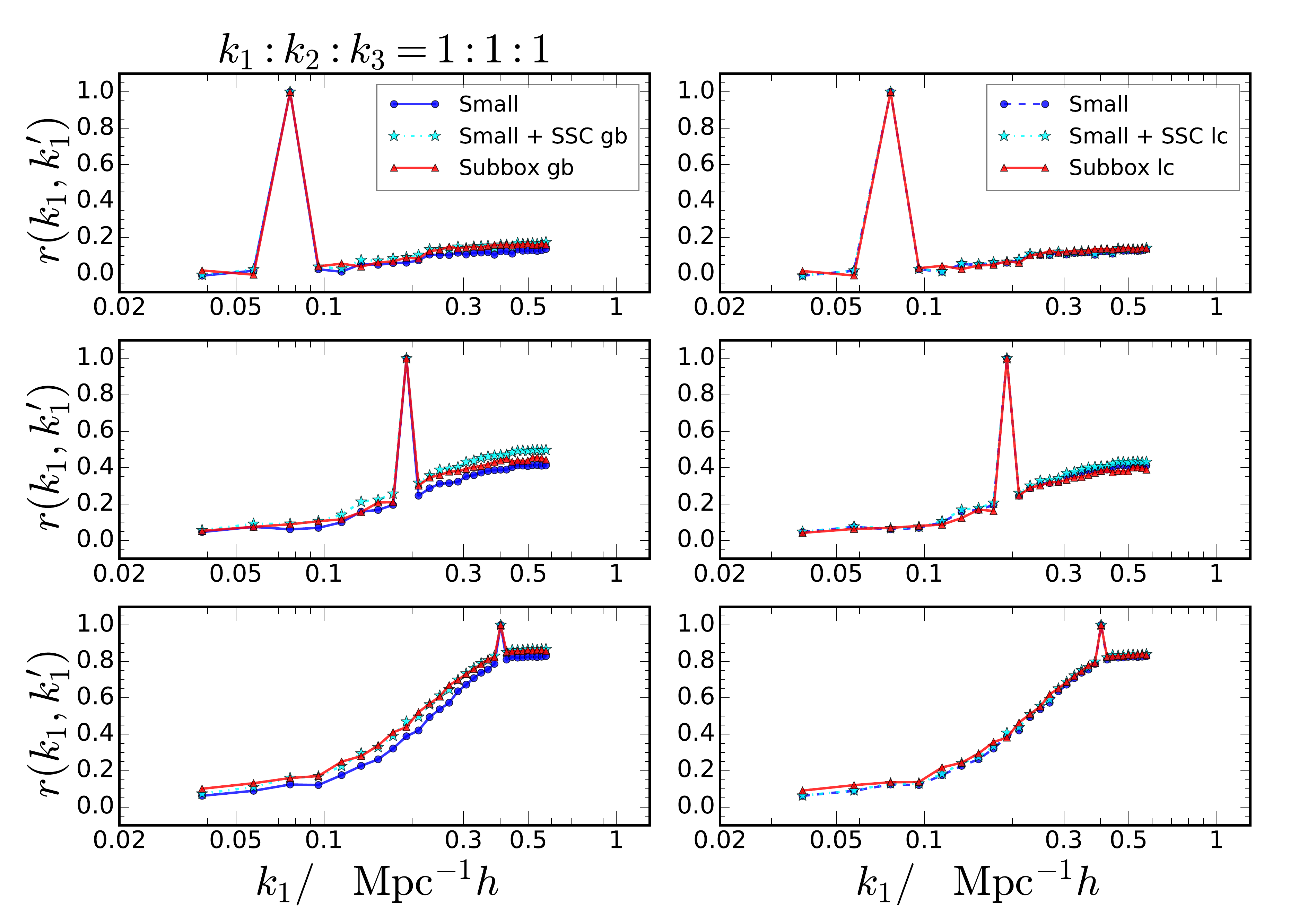}
\caption{ The cross correlation coefficient $r(k_i,k_j)$ for the equilateral triangle configurations at $z=0$. In each plot, $k_i$ is fixed to be 0.077, 0.19, and 0.77 $\hOMpc$ respectively (from the top to bottom row), and it is plotted as a function of $k_j$.    The left panels are for the global mean case, while the right ones are for the local mean.  The numerical results from the small box (blue, circles) and the subbox setup (red, triangles) are compared with the supersample covariance prediction (cyan, stars).    } 
\label{fig:corr_coef_small_subbox_111_SSC}
\end{center}
\end{figure*}

\begin{figure*}[!htb]
\begin{center}
\includegraphics[width=0.9\linewidth]{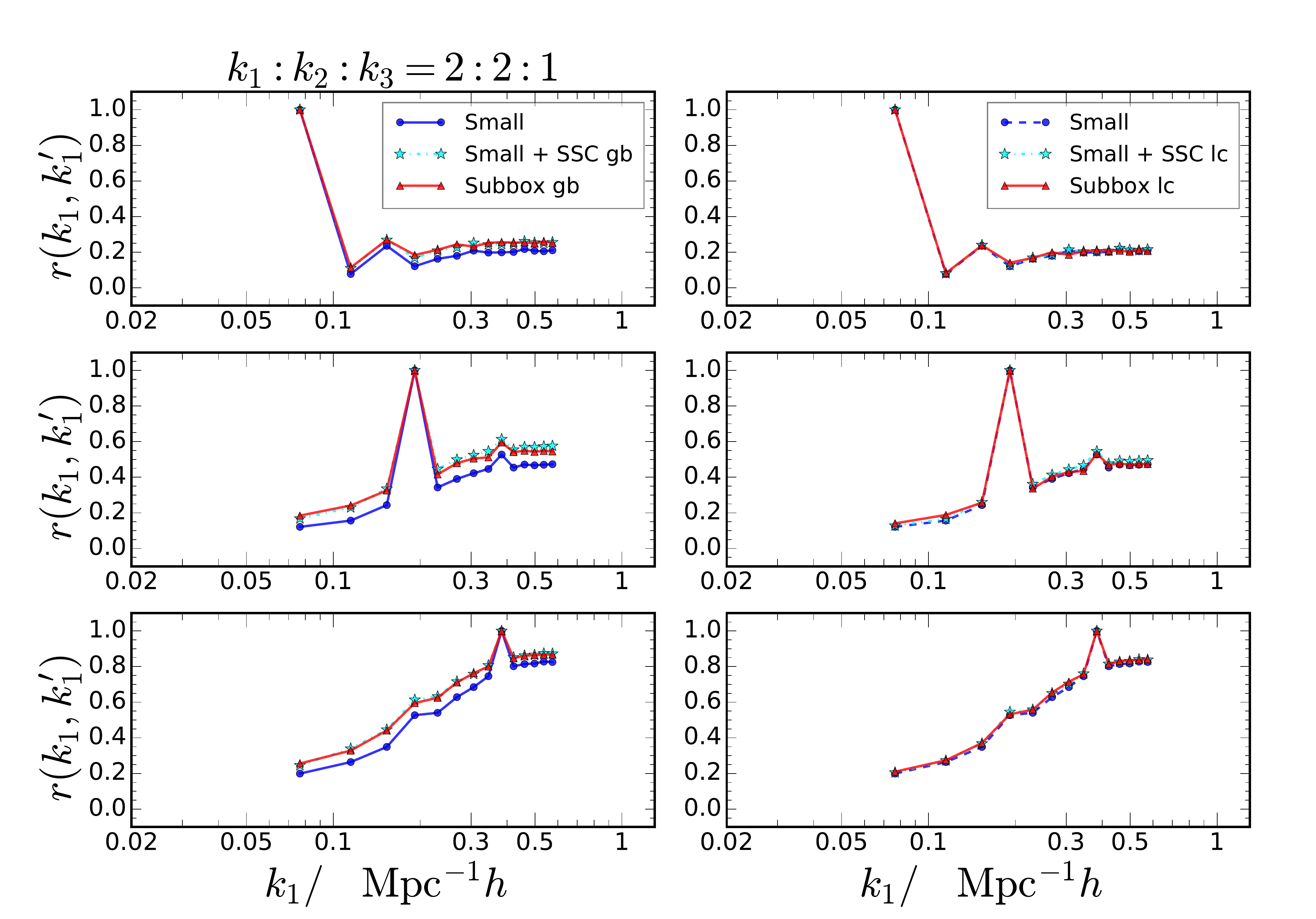}
\caption{  Similar to Fig.~\ref{fig:corr_coef_small_subbox_111_SSC} except the for isosceles triangle of the shape $k_1: k_2 :k_3 = 2:2:1 $.  } 
\label{fig:corr_coef_small_subbox_221_SSC}
\end{center}
\end{figure*}

\section{ The supersample covariances: predictions and measurements }
\label{sec:predictions_measurements}

In this section, we compute the supersample covariance for the bispectrum and the cross covariance between the power spectrum and the bispectrum. The predictions are then confronted with the measurements from simulation.

Before going to the numerical results, we first compare the supersample covariance contributions with the Gaussian covariance, which is valid in the low $k$ regime.   The Gaussian bispectrum covariance reads \cite{Scoccimarro:2003wn}
\beq
\label{eq:BCov_Gaussian} 
C^B_{\rm G} =  \frac{( 2 \pi)^3  k_{\rm F}^3 }{V_{\triangle} }  \delta_{k_1k_2k_3, k_1'k_2'k_3'  } s_{123}   P(k_1) P(k_2) P(k_3) ,
\eeq
where  $\delta_{k_1k_2k_3, k_1'k_2'k_3'  }$  is non-vanishing only if the shape of the triangle  $ k_1k_2k_3 $ is the same as that of $ k_1'k_2'k_3' $. The symmetry factor $s_{123} $ is equal to 1, 2, and 6 for scalene, isosceles, and equilateral triangle respectively. In Eq.~\eqref{eq:BCov_Gaussian}, $P(k) $ can be the linear power spectrum or the nonlinear one, in this case, it effectively resums part of the higher-order contributions.  For the cross covariance between the power spectrum and the bispectrum, the Gaussian contribution vanishes as it is a 5-point correlator.

To facilitate the comparison with the simulation results, we consider a cubic survey window
\beq
W_{\rm cubic}(\bm{x}) =  \begin{cases} 
      1   & \mathrm{if} \,  0 \leq x_i \leq L,  \\
      0    &  \mathrm{otherwise} . 
   \end{cases}
\eeq
Its Fourier transform reads
\beq
W_{\rm cubic}( \bm{k} ) = V e^{- i \bm{k} \cdot \bm{L} / 2  }  \prod_{j=1}^3 \mathrm{sinc} \frac{ k_j L }{2}  .
\eeq
In the supersample covariance formula, it is the RMS variance $\sigma_W^2$ that matters.  If a spherical tophat window is used instead, we find that by matching the survey volume with the mapping $R_{\rm TH} =  [3 /(4 \pi )]^{1/3} L $, the variance computed with either window agrees with each other within at least 3\% for the box sizes we consider here. Thus, this provides a convenient way to map our results to the tophat window case.

In Fig.~\ref{fig:CB_SSC_response_111} we compare the  supersample covariance with the Gaussian one for a range of box sizes: 600, 1000,  and 3000 $\MpcOh$ (at $z=0$, the corresponding $\sigma_W^2 $ are $9.8 \times 10^{-5}$,  $1.7 \times 10^{-5}$, and  $3.3 \times 10^{-7}$, respectively). We have compared the supersample covariance predictions using the perturbation theory and the halo model response function.  Current surveys such as DES have a survey volume close to the volume of the 1000 $\MpcOh$ box, and the future survey Euclid will have a survey volume  near that of the 3000 $\MpcOh$ box. In this plot, we have kept the bin width fixed at $\Delta k = 2 k_{\rm F} $; thus, the number of modes in each configuration bin and the Gaussian covariance is constant for different box sizes.  Because the bispectrum Gaussian covariance is very subdominant beyond the mildly nonlinear regime relative to the small scale non-Gaussian covariance \cite{Chan:2016ehg}, the supersample covariance is expected to only give a small contribution to the overall bispectrum covariance budget. We will quantify this using the simulation results below.


We now compare the covariances measured from the periodic box and subbox setups. The simulations used in this work are from the DEUS project \cite{DEUS_FUR_paper,Rasera:2013xfa,Blot:2014pga}.  A flat $\Lambda$CDM model with the WMAP7 cosmological parameters \cite{2007ApJS..170..377S} is adopted for these simulations. In particular, $h=0.72$, $\Omega_{\rm m} = 0.257$, $n_{\rm s } = 0.963$, and $\sigma_{8 } = 0.801 $. The Zel'dovich approximation is used to generate the Gaussian initial conditions  at $z_i = 105$.  The transfer function is computed with {\tt CAMB} \cite{CAMB}.  The simulations are evolved using the adaptive mesh refinement solver {\tt RAMSES} \cite{2002A&A...385..337T}. We will only consider dark matter simulation results at $z=0$.

The periodic simulations are the small set used in \cite{Chan:2016ehg}.  In each periodic simulation, there are $256^3$ particles in a cubic box of size $L=656.25\MpcOh $. There are altogether 4096 realizations. For the subbox setup, we use a gigantic simulation of box size 21 Gpc$ \, h^{-1}$ with $8192^3 $ particles from the DEUS full universe run. The gigantic box is divided into cubic subboxes of size $656.25 \MpcOh$. There are altogether 32768 subboxes and we use 4096 of them.  

To facilitate the comparison with the bispectrum results later on, we show the power spectrum results here as well.  Similar to \cite{Takahashi:2009ty,Li:2014sga}, we find that the power spectrum measurement is biased low due to the window function convolution in the range of scales we consider.  The bias depends on the shape of the power spectrum and the size of the window function.  For our case, it is most substantial in the range from $k \sim  0.01$ to 0.1 $\hOMpc$. Ref.~\cite{Li:2014sga} scaled the value of the subbox case to match the periodic box measurement, here we do not apply any correction as we find that this helps little for the case of bispectrum.  We plot the diagonal elements of the power spectrum covariance ratio between the subbox results and the small box results in Fig.~\ref{fig:PkCovDiag_ratio_subbox_small_SSC_z0}. At $k=0.5 \hOMpc$, the supersample covariance correction to the small box results is about 60\% for local mean and  250\% for the global mean. We will see that the effect is substantially smaller for the bispectrum, and it is of similar order of magnitude for the cross covariance.

For the bispectrum, the expectation value of Eq.~\eqref{eq:B_w_form1} reads
\begin{align}
\langle  & \hat{B}_W (k_1,k_2,k_3) \rangle  = \frac{1}{V  V_{\triangle}} \int_{k_1} d^3 p_1  \int_{k_2} d^3 p_2  \int_{k_3} d^3 p_3  \Ddel(\bm{p}_{123} ) \nn \\
& \quad  \times \prod_{i=1}^3 \int \frac{ d^3 q_i}{ (2 \pi)^3 } W(\bm{p}_i - \bm{q}_i ) (2 \pi)^3 \Ddel( \bm{q}_{123} ) B( q_1,q_2,q_3). 
\end{align}
The first line simply averages the triangles within the configuration bin, and hence any bias is expected to arise from the smearing effect by the window in the second line. The window function satisfies
\beq
\label{eq:W3_convolution_intg}
\frac{1}{V}  \prod_{i=1}^3 \int \frac{ d^3 q_i}{ (2 \pi)^3 } W(\bm{p}_i - \bm{q}_i ) (2 \pi)^3 \Ddel( \bm{q}_{123} ) = 1,
\eeq
where we have used $\bm{p}_{123} = \bm{0} $ and Eq.~\eqref{eq:Wn_identity_real}. Hence we can interpret that the window convolution simply results in a weighted mean of the bispectrum. For a more extended window, the value in Eq.~\eqref{eq:W3_convolution_intg} is smaller, e.g.~for Gaussian window, it is  $1/ \sqrt{27} $ instead of 1, and so the windowed bispectrum is more biased relative to the underlying one. Except for the first bin, which is biased low, the subsequent bins are biased high (by roughly  2\%) and the bias decreases as $k$ increases.   Apart from the bias in the amplitude, the subbox measurements also exhibit wiggles, which are strongest in the low $k$ regime and decrease as $k$ increases.

In Fig.~\ref{fig:diag_cov_mat_small_subbox_111_SSC}, we show the diagonal element of the bispectrum covariance for the equilateral triangle configuration obtained from the small box and the subbox setups. In the left panel, the covariance is normalized with respect to the Gaussian covariance. For both the global and local mean cases, the supersample covariance correction is small relative to the small scale covariance, which can be studied using a periodic setup. To see the differences more clearly, we show the ratio between the subbox covariance and the small box covariance in the right panel.  We see clearly that there are wiggles in the subbox covariance due to the convolution with the cubic window. 
Again the effect of the supersample covariance is small, up to $k \sim 0.5 \hOMpc $, the enhancement for the covariance is about 30 \% for the global mean and about 5\% for the local mean. The ratio is roughly an order of magnitude smaller than that for the power spectrum covariance (Fig.~\ref{fig:PkCovDiag_ratio_subbox_small_SSC_z0}).   We show the ratio between the subbox setup  and the periodic box results for the isosceles triangle configurations $ k_1:k_2:k_3=2:2:1 $ as a function of $k_1$  in Fig.~\ref{fig:diag_cov_mat_small_subbox_221_SSC}. Clearly they are qualitatively similar to the equilateral triangle case.

We also show the supersample covariance prediction given by
\beq
C_{\rm Small+SSC } = C_{\rm Small} +  C_{\rm SSC}, 
\eeq
where $  C_{\rm Small} $ is the covariance measured from the small set and $  C_{\rm SSC} $ is obtained with Eq.~\eqref{eq:CB_SSC}  using the halo model response function. We find that for both the equilateral triangle and isosceles triangle configurations, besides the convolution due to the survey window, the  supersample covariance prediction agrees with the subbox results well up to $k\sim 0.5 \hOMpc $.   However, the halo model predictions overpredict the effect for larger $k$; e.g.~at $k=1\hOMpc$, it is overpredicted by 20\% for local mean and 170\% for the  global mean.

The cross correlation coefficient $r$ is generally defined as 
\begin{align}
  & r(k_1,k_2,k_3, k_1',k_2',k_3') \nn \\
= & \frac{ C(k_1,k_2,k_3, k_1',k_2',k_3') }{\sqrt{  C(k_1,k_2,k_3, k_1,k_2,k_3)   C( k_1',k_2',k_3', k_1',k_2',k_3')    }   }. 
\end{align}
By definition it is equal to 1 for the diagonal term.  In Fig.~\ref{fig:corr_coef_small_subbox_111_SSC}, we plot $r$ for the equilateral triangle configurations. In these plots, one of the equilateral triangle is fixed to be of the size 0.077, 0.19, and 0.77 $\hOMpc$ respectively. Again, we find that the difference between small box and the subbox setup is small for the global mean case, and it is negligible for the local mean scenario. Furthermore, the supersample covariance prediction gives decent agreement with the subbox results.  In Fig.~\ref{fig:corr_coef_small_subbox_221_SSC}, similar results are shown except it is for the isosceles triangle configuration $k_1:k_2:k_3 = 2:2:1 $. Here for the isosceles triangle $k_1':k_2':k_3'=2:2:1$, $k_1'$ is set to   0.077, 0.19, and 0.77 $\hOMpc$ respectively and it is plotted as a function of $k_1$. Again the results are similar to the equilateral triangle case.

\begin{figure*}
\begin{center}
\includegraphics[width=0.9\textwidth]{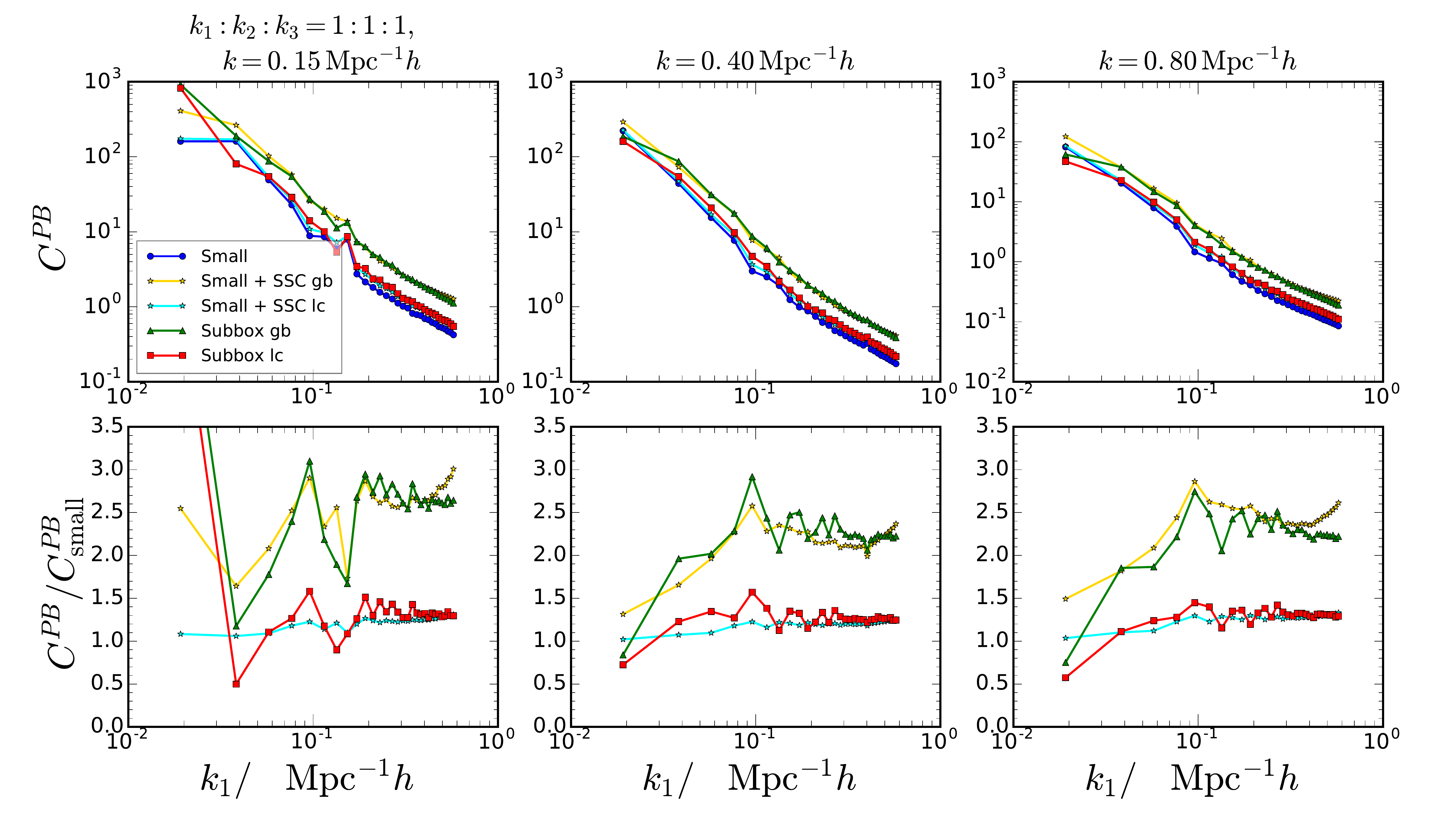}
\caption{ The cross covariance between the matter power spectrum and bispectrum at $z=0$. The equilateral triangle configurations are used for the bispectrum and the Fourier mode of the power spectrum $k$ are chosen to be 0.15, 0.40 and 0.80 $\hOMpc$ (from left to right). In the lower panels, the covariances are normalized with respect to the small box results.   The results from the small box (blue, circles) and subbox results for the global mean (green, triangles) and the local mean cases (red, squares) together with the supersample covariance prediction for the global mean (yellow, stars) and local mean (cyan, stars) cases are compared. }
\label{fig:CPB_Small_subbox_SSC_111}
\end{center}
\end{figure*}

\begin{figure*}
\begin{center}
\includegraphics[width=0.9\textwidth]{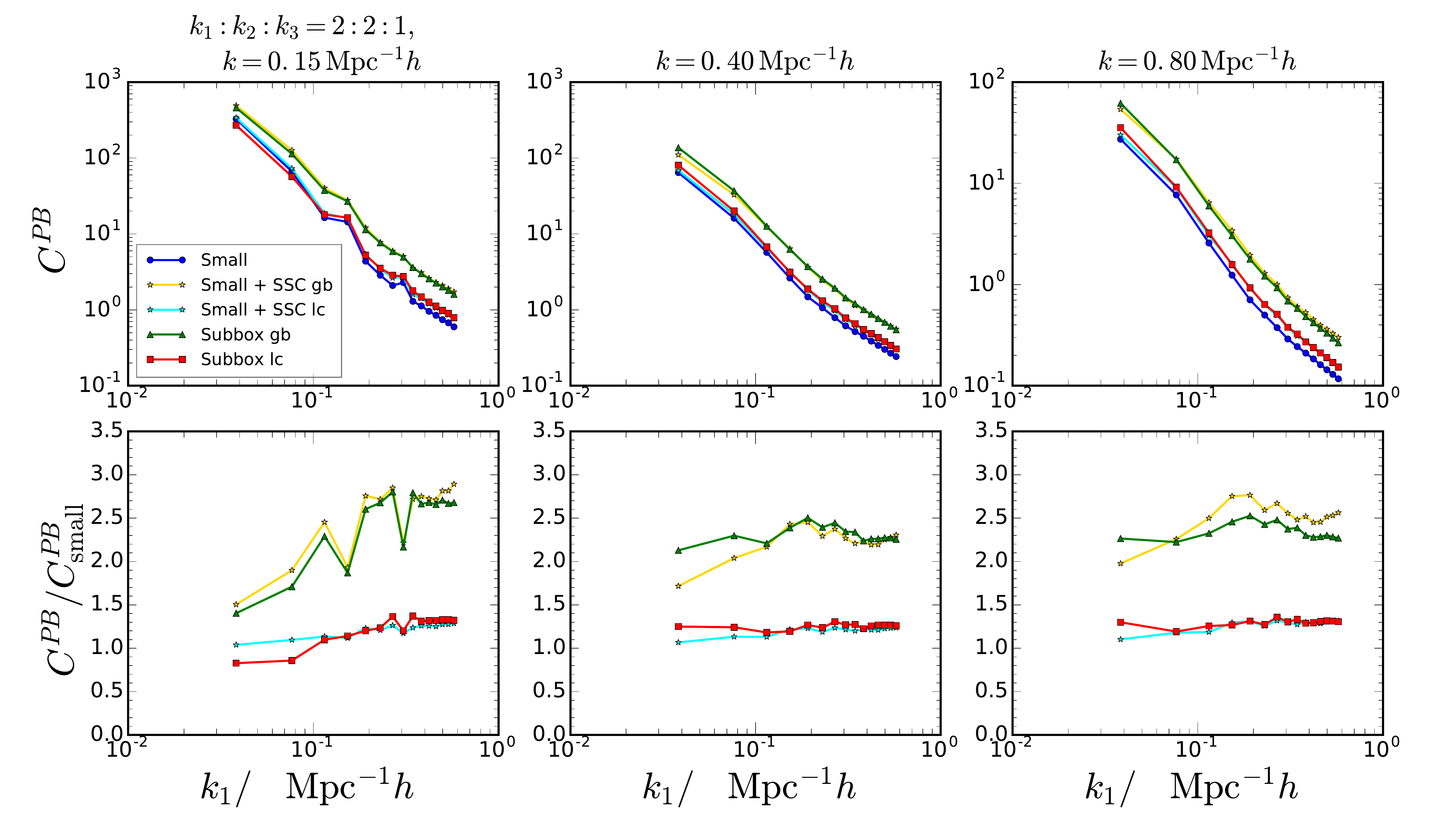}
\caption{ Similar to Fig.~\ref{fig:CPB_Small_subbox_SSC_111} except for the bispectrum shape $k_1 : k_2 : k_3 = 2:2:1 $.   }
\label{fig:CPB_Small_subbox_SSC_221}
\end{center}
\end{figure*}

We now look at the cross covariance between the matter power spectrum and the bispectrum. In Figs.~\ref{fig:CPB_Small_subbox_SSC_111} and \ref{fig:CPB_Small_subbox_SSC_221} we show the cross covariance, $C^{PB}(k ;k_{1},k_{2},k_{3}) $ with the Fourier mode of the power spectrum $k$ fixed to be 0.15, 0.40 and 0.80 $\hOMpc $ respectively. In Fig.~\ref{fig:CPB_Small_subbox_SSC_111} the bispectrum is chosen to be the equilateral triangle configurations, while they are  isosceles triangles $ k_1:k_2:k_3=2:2:1 $ in Fig.~\ref{fig:CPB_Small_subbox_SSC_221}.   By  comparing with Fig.~\ref{fig:diag_cov_mat_small_subbox_111_SSC} and \ref{fig:diag_cov_mat_small_subbox_221_SSC}, we find that the fractional difference between the small box and subbox setups is larger than the bispectrum covariance alone.   This is mainly because the small scale non-Gaussian covariance of the bispectrum is bigger than the power spectrum one, and thus the supersample  covariance contribution is relatively small for the bispectrum.  The order of magnitude is similar to that for the power spectrum. We also show the supersample covariance prediction Eq.~\eqref{eq:SSC_PB_prediction} with the response functions computed using the halo model prescriptions. Similar to the bispectrum covariance case, the prediction gives good agreement with the numerical results for both the global mean and local mean scenarios. However, we also note that for large $k$ (e.g.~$k=0.8\hOMpc$) the theory  is clearly larger than the measurement for the global mean case.


In the comparison with the numerical results, we have only considered the supersample covariance contribution due to the long-wavelength density perturbutions. In Appendix ~\ref{sec:SSC_tides},  we work out the tidal perturbation contribution to the supersample covariance. Although the magnitude of the tidal response function and the corresponding variances [$S_{ijmn}$, defined in Eq.~\eqref{eq:Sijmn}] are comparable to their density counterparts, the net tidal supersample covariance contribution is smaller than the density one by a few orders of magnitude for the following reasons.  In configuration space, the bispectrum depends only on the shape of the triangle, and not on its orientations. The tidal response function is anisotropic, and so after averaging over the orientations of the triangle its contribution is significantly reduced. See Appendix~\ref{sec:SSC_tides} for more details.  However, in redshift space, the bispectrum is anisotropic, and so this contribution could be potentially larger. We leave the thorough investigation of this issue to future work.

Before closing this section, we would like to extrapolate the results here to estimate the relative importance of various covariance contributions for future surveys like Euclid.  Here we take the survey volume of Euclid to be equivalent to a cubic box of size 4000 $\MpcOh$ and the mean redshift to be 1.2.  In the perturbative regime,  $ [ (\partial B / \partial \delta_{\rm b} )|_{ \delta_{\rm b }=0}]^2 \propto D^8 $ and  $ (\partial P / \partial \delta_{\rm b} )|_{ \delta_{\rm b }=0}  (\partial B / \partial \delta_{\rm b} )|_{ \delta_{\rm b }=0}  \propto D^6 $. On the other hand, the leading perturbative non-Gaussian corrections for the covariance of $B$ and $P$-$B$ scale as $D^8$  and $D^6$ respectively  \cite{Chan:2016ehg}. Because both contributions have the same perturbative time dependence, to translate the small box results at $z=0$ to the Euclid setting, we only need to compare $\sigma_W^2(z) $ with $V^{-1}$. Note that the small scale covariance scaling with volume has been checked in \cite{Chan:2016ehg} (see also \cite{Mohammed:2016sre}) using simulations of different sizes, and it was found to be in good agreement with the numerical results. Hence the supersample covariance for $B$ or $P$-$B$ is reduced by a factor of 2320 in the Euclid setting, while the small scale covariance is only suppressed by a factor of 227. The relative importance of the supersample covariance compared to the small scale one would be downgraded by a factor of 10 in Fig.~\ref{fig:diag_cov_mat_small_subbox_111_SSC},  \ref{fig:diag_cov_mat_small_subbox_221_SSC}, \ref{fig:corr_coef_small_subbox_111_SSC}, \ref{fig:corr_coef_small_subbox_221_SSC}, \ref{fig:CPB_Small_subbox_SSC_111}, and \ref{fig:CPB_Small_subbox_SSC_221}.

In this section, we have quantified the magnitudes of the supersample covariance contributions by comparing the numerical results obtained with the periodic box and subbox setups. We have also tested the supersample covariance prediction derived in the previous sections  and found that it  agrees well with the numerical results.  Although we have only explicitly shown the two types of triangle configurations, the results are qualitatively similar for other configurations. The numerical results and the predictions are validated by their good  agreement with each other.  For example, the transients induced by the Zel'dovich approximation initial condition for the bispectrum \cite{McCullagh:2015oga} do not seem to be an issue here.

\section{Conclusions }
\label{sec:conclusions}

In the current and future large-scale structure surveys, the quality of the data is expected to keep on increasing. At the same time, to extract cosmological information from such high-fidelity data, the theoretical modeling precision of various systematics also needs to increase. One of the potential systematics is the supersample covariance.  In the presence of the window function, the long mode with wavelengths larger than the survey window can modulate the small scales and cause large covariance inside the survey window \cite{Hamilton:2005dx}. The supersample covariance cannot be studied using the standard periodic simulation setup.  The window function effectively broadens the wave vectors, while in the standard periodic simulation setup the wave vectors are sharp.  This broadening can be captured by dividing a gigantic simulation into many subboxes.  In \cite{Takada:2013bfn}, the power spectrum supersample covariance was formulated using the response function approach.  The power spectrum supersample covariance has been recognized an important source of covariance on small scales.

In this paper we studied the supersample covariance contribution to the bispectrum covariance and cross covariance between the power spectrum and the bispectrum. In terms of the response function, we derived the supersample covariance for the bispectrum covariance [Eq.~\eqref{eq:CB_SSC}] and for the cross covariance [Eq.~\eqref{eq:SSC_PB_prediction}]. We also computed the bispectrum response function using the standard perturbation theory [Eq.~\eqref{eq:Bk_response_SPT_global}] and the halo model [Eq.~\eqref{eq:BHM_response}].  Besides the density, we also derived the bispectrum supersample covariance due to the tide [Eq.~\eqref{eq:SSC_tide_appendix}] and the bispectrum response function to the tide [Eq.~\eqref{eq:response_tide}] using SPT. However, we found that the tide contribution to the supersample covariance is a few orders of magnitude smaller than the density one because the bispectrum in configuration space is isotropic while the tide response function is anisotropic.

We quantified the magnitudes of the supersample covariance using numerical measurements with the periodic box and subbox setups. The effects are small for the bispectrum covariance with the global mean, and for the local mean case it is negligible (Figs.~\ref{fig:diag_cov_mat_small_subbox_111_SSC} -- \ref{fig:corr_coef_small_subbox_221_SSC}).  Relative to the small scale covariance, the magnitude of the supersample covariance is roughly an order of magnitude smaller than the power spectrum case.   This is because the small scale covariance for the bispectrum is much more significant than in the power spectrum case, e.g.~by comparing with the Gaussian covariance \cite{Chan:2016ehg}, thus the supersample covariance contribution is dwarfed.  For the cross covariance, the effect is larger, and closer to that of the power spectrum covariance (Figs.~\ref{fig:CPB_Small_subbox_SSC_111} and \ref{fig:CPB_Small_subbox_SSC_221}). Thus in the combined analysis of the power spectrum and the bispectrum, the supersample covariance may not be negligible. However, in galaxy surveys, a local mean is used, and hence the supersample covariance is still a small correction to the total covariance budget. We can also directly use the halo model supersample covariance because we find that it works reasonably well and the supersample covariance is a small correction anyway. 

Ref.~\cite{Chan:2016ehg} found that the small scale non-Gaussian covariance is much more significant for the bispectrum than for power spectrum, and speculated that the small scale covariance is even more serious for the higher order correlators. Along a similar vein, we surmise that the supersample covariance is even  more subdominant for the higher order correlators, and hence negligible.

Our work makes it clear that for the bispectrum covariance and the cross covariance, the small scale covariance is the dominant source, at least up to $k \sim 1 \hOMpc$. For the bispectrum this is probably the highest scale we can hope to model. Thankfully, the small scale non-Gaussian covariance can be studied using the standard periodic setup with small box size, which is much more accessible in terms of computational resources.  On the other hand, there have been few efforts so far to model the small scale covariance \cite{Sefusatti:2006pa,Chan:2016ehg}. The perturbative approach only improves over the Gaussian covariance in the mildly nonlinear regime \cite{Chan:2016ehg}. To extend the perturbative calculation to higher $k$, one possibility is to model the bispectrum covariance using the halo model. A useful way of organizing the computation is to expand the covariance in terms of the connected correlators \cite{Sefusatti:2006pa,KayoTakadaJain_2013,Chan:2016ehg}, which in turn are computed using the halo model.

\section*{Acknowledgments} 

We thank Linda Blot for discussions on the DEUS simulations and the anonymous referee for constructive comments that improves the draft.  We are grateful for  the members of the DEUS Consortium\footnote{\url{http://www.deus-consortium.org}} for sharing the data with us. K.C.C. acknowledges the support from the Spanish Ministerio de Economia y Competitividad grant ESP2013-48274-C3-1-P and the Juan de la Cierva fellowship. J.N. is supported by Fondecyt Grant 1171466.

\appendix

\section{ Beat coupling for the bispectrum    } 
\label{sec:beat_coupling_derivation}

The effects of the long mode on the small scale power spectrum covariance were first computed in Ref.~\cite{Hamilton:2005dx}, using both perturbation theory and a nonlinear model, the hyper extended perturbation theory. The authors coined the term ``beat coupling'' to refer to the covariance induced by the long mode on the modes within the survey window, i.e.~supersample covariance.  In this paper we follow Refs.~\cite{Takada:2013bfn,Li:2014sga} to use beat coupling to refer only to the perturbative part of the supersample covariance.

In this Appendix, we compare the beat-coupling contribution computed with the approach of  Hamilton et al (HRS) \cite{Hamilton:2005dx} against the full perturbative computation done in the main text. Let us start with the case of the power spectrum.  It was pointed out in \cite{Hamilton:2005dx} that at the tree level, an additional diagram sourcing the trispectrum due to the long mode arises when the window function is present. For the case of dark matter, this additional tree level trispectrum reads
\beq
T( \bm{k},\bm{k}',\bm{\epsilon} ) = 16 P(k) P(k') P(\epsilon)  F_2( -\bm{k}_1, \bm{\epsilon} )F_2( -\bm{k}_1', - \bm{\epsilon} ), 
\eeq
where $\bm{\epsilon}$ represents the long mode. Replacing the $F_2$ kernel by the one spherically averaged over the solid angle of the long mode
\beq
\label{eq:F2_sph_average}
\int \frac{ d \Omega_\epsilon }{ 4 \pi } F_2( \bm{k}, \bm{\epsilon}  ) = \frac{17}{21},
\eeq
we get 
\beq
\bar{T}( k, k', \epsilon ) = \frac{4624 }{ 441 } P(k)P(k') P(\epsilon).   
\eeq
Similar to calculations in Sec.~\ref{sec:SSC_Bk_derivation}, integration of long mode $\epsilon$ yields the beat coupling covariance
\beq
\label{eq:CP_beatcoupling}
C_{\rm BC}^P(k,k') =  \frac{4624 }{ 441 } P(k)P(k') \sigma_W^2. 
\eeq

The result is different from that in Sec.~\ref{sec:response_SPT_longshortcoupling} because the dilation effect is neglected and we have used a different angular averaged $F_2$: $\langle F_2(\bm{k}- \bm{\epsilon}, \bm{\epsilon}) \rangle_{\Omega_\epsilon} = 13/21$. Compared to the results obtained in  Sec.~\ref{sec:response_SPT_longshortcoupling}, Eq.~\eqref{eq:CP_beatcoupling} overestimates the covariance by about 30\%  at $k = 0.2 \hOMpc $.

Next we consider the bispectrum covariance using the approach of \cite{Hamilton:2005dx} and check how this compares with our result in Sec.~\ref{sec:response_SPT_longshortcoupling}.  The bispectrum covariance is sourced by the 6-point function, including the connected and disconnected diagrams.  The disconnected contributions include the Gaussian covariance which has three disconnected components and  non-Gaussian contributions, which have two disconnected parts.  Ref.~\cite{Chan:2016ehg} computed the tree level disconnected terms using perturbation theory, and they also estimated the tree level connected contribution and found that they are sub-leading relative to the disconnected non-Gaussian contributions in the perturbative regime.

\begin{figure}[!tb]
\begin{center}
\includegraphics[width=\linewidth]{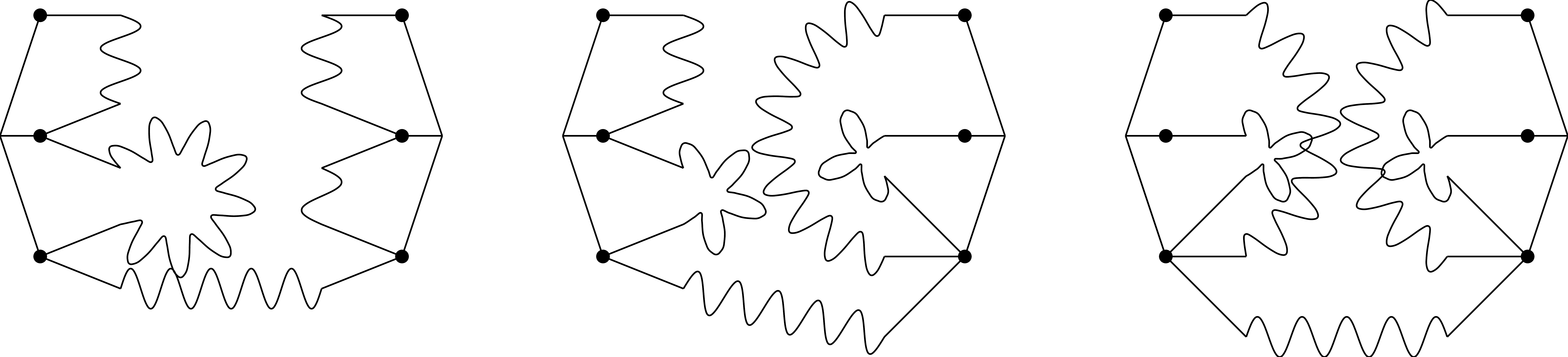}
\caption{ A diagrammatic representation of the beat coupling terms  for the bispectrum. The three dots on either side represent the three $\delta$'s in the bispectrum estimator. The legs from each dot denote the $F_n$ kernel, while the wavy line represents the power spectrum.   }
\label{fig:Covmat_Bk_BeatCoupling}
\end{center}
\vspace{-0.5cm}
\end{figure}

In the presence of the window function, three additional connected diagrams  arise.  In Fig.~\ref{fig:Covmat_Bk_BeatCoupling}, we show the diagrammatic representation of these beat coupling terms for the bispectrum (cf.~the ``normal'' tree level diagrams in Fig.~9 of ~\cite{Chan:2016ehg}).   A set of three dots on either side of the diagram  denotes the bispectrum estimator. The solid lines branching from each dot represent the $F_n$  kernel.  The wavy line denotes the power spectrum.   The important feature is that the two sides of the bispectrum estimator are linked by one and only one  leg from the $F_n$ with $n>1$. We also see this feature in the beat coupling term for the power spectrum \cite{Hamilton:2005dx}.  In the absence of the window functions, because of the closed triangle condition, the sum of the wave vectors vanish on both sides of the bispectrum estimator. Hence the contribution due to these diagrams vanishes because $P(0)=0$. That is why these diagrams are not present in \cite{Chan:2016ehg}.  The story changes when  the window function is present, and  now the long mode with finite wave number connects the two sides .

The beat coupling contribution to the 6-point function is
\begin{align}
  \label{eq:YBC1}
  Y_{\rm BC 1}&  ( k_1,k_2,k_3, k_1',k_2',k_3',\bm{\epsilon}) \nn \\
   = & 16  P(k_1) P(k_3) F_2(\bm{k}_1,\bm{k_3}) P(k_1') P(k_3')F_2(\bm{k}_1',\bm{k_3}')    \nn \\
  & \times   F_2(\bm{k}_3 , \bm{\epsilon} ) F_2(\bm{k}_3', - \bm{\epsilon} ) P(\epsilon)    + \,  35 \, \mathrm{cyc},  \\
  \label{eq:YBC2}
 Y_{\rm BC 2}&(  k_1,k_2,k_3, k_1',k_2',k_3',\bm{\epsilon}) \nn \\
   = &   24 P(k_1) P(k_3)   F_2(  \bm{k}_1, \bm{k_3} )   F_2( \bm{k}_3 , \bm{\epsilon})   P(k_1') P(k_2') \nn \\
    &  \times F_3(\bm{k}_1', \bm{k}_2', \bm{\epsilon} )  P(\epsilon)    +  17 \, \mathrm{cyc} + ( \bm{k} \leftrightarrow \bm{k}' ),   \\
  \label{eq:YBC3}
 Y_{\rm BC 3}& ( k_1,k_2,k_3, k_1', k_2',k_3',\bm{\epsilon}) \nn \\
 = &    36 P(k_1) P(k_2) F_3( \bm{k}_1, \bm{k}_2, -\bm{\epsilon} )  P(k_1')  P(k_2')   \nn \\
   &  \times F_3( \bm{k}_1',  \bm{k}_2',  \bm{\epsilon} )       P(\epsilon)    \,   +  8 \, \mathrm{cyc}.  
\end{align}

\begin{figure}[t]
\begin{center}
\includegraphics[width=\linewidth]{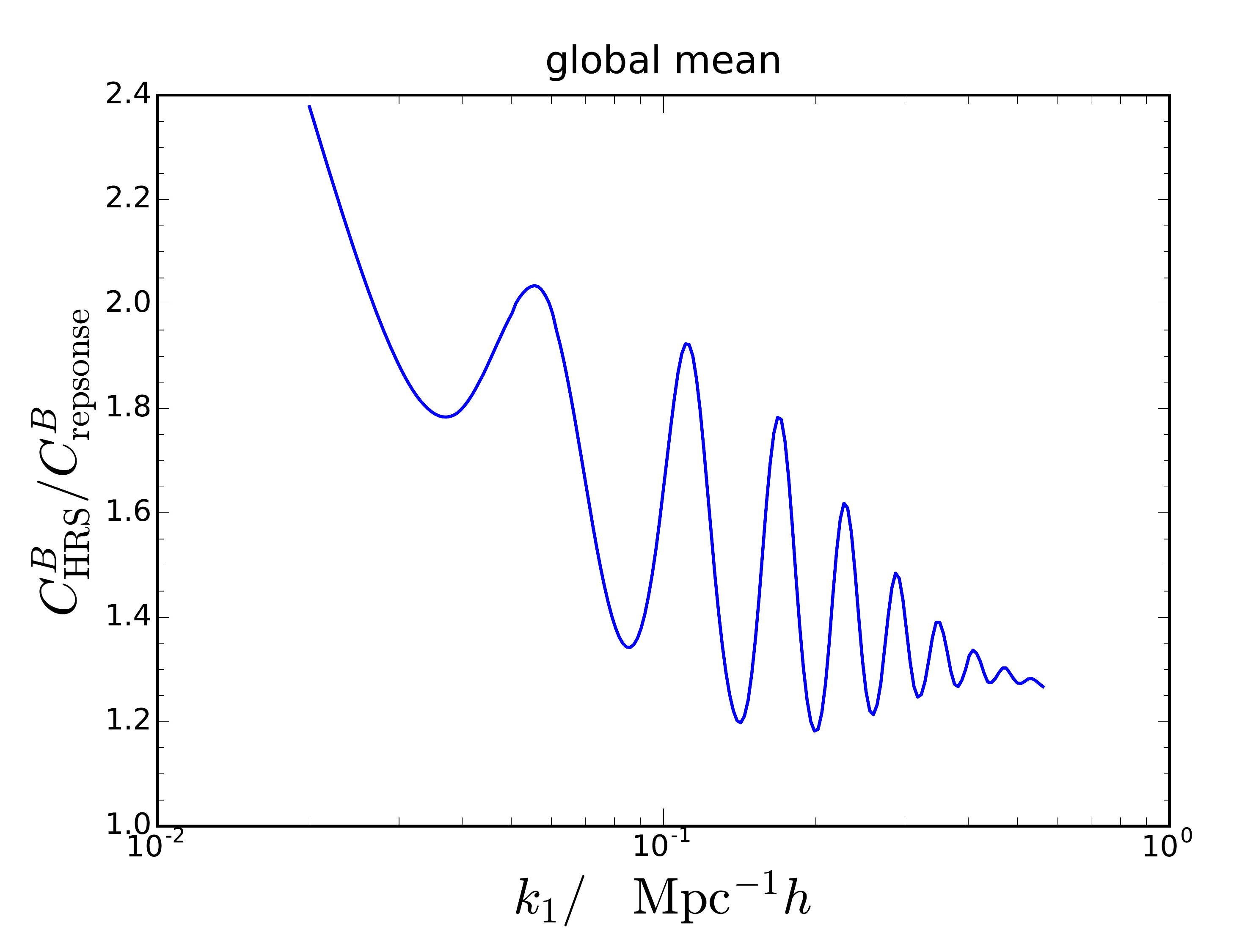}
\caption{  The ratio between the diagonal elements of the beat-coupling bispectrum covariance obtained using the method of HRS \cite{Hamilton:2005dx} and that from the response approach described in the main text. The global mean and the equilateral triangle configuration are used.  }
\label{fig:CB_HRS_SSC_ratio_11}
\end{center}
\end{figure}

To simplify Eqs.~\eqref{eq:YBC1}-\eqref{eq:YBC3}, we replace the $ F_2 $ with the long mode by Eq.~\eqref{eq:F2_sph_average}, and $ F_3 $ with the long mode by the spherical average 
\begin{align}
\int \frac{ d \Omega_\epsilon }{ 4 \pi } F_3( \bm{k}_1, \bm{k}_2, \bm{\epsilon}  ) = \frac{1}{378} [ 277  F_2( \bm{k}_1,\bm{k}_2 ) 
       +   5 G_2(  \bm{k}_1,\bm{k}_2 )  ] .  
\end{align}
Then the  6-point functions reduce to
\begin{align}
  \label{eq:YBC1_sphave} 
  & Y_{\rm BC 1 }( k_1,k_2,k_3, k_1',k_2',k_3', \epsilon ) 
    =  \frac{4624  }{ 441 }   B_{\rm m} B_{\rm m}'  P(\epsilon) ,  \\
  \label{eq:YBC2_sphave} 
  &  Y_{\rm BC 2 }( k_1,k_2,k_3, k_1',k_2',k_3', \epsilon )   \nn \\
  & \quad    =   \frac{ 68 }{1323  } \bigg[   277  B_{\rm m} B_{\rm m}'  
    +  \frac{5}{2} \Big(  B_{\rm m} B_\theta'  +  B_{\rm m}' B_\theta    \Big)     \bigg]  P(\epsilon) , \\
    \label{eq:YBC3_sphave} 
     &  Y_{\rm BC 3 }( k_1,k_2,k_3, k_1',k_2',k_3', \epsilon )   \nn \\
   & \quad  =  \frac{1}{15876}  ( 277 B_{\rm m} + 5 B_\theta) ( 277 B_{\rm m}' + 5 B_\theta')   P(\epsilon)  ,
\end{align}
where $B_{\rm m}' = B_{\rm m}( k_1', k_2', k_3' ) $ and etc, and  $B_{\theta} $ denotes
\beq
B_\theta(k_1,k_2,k_3) = 2 G_2( \bm{k}_1, \bm{k}_2 ) P(k_1) P(k_2) + 2 \, \mathrm{cyc.}
\eeq

The long mode  $\bm{q}_{123} $ in Eq.~\eqref{eq:BB_expectation} is represented by $\bm{\epsilon } $, and we can plug Eq.~\eqref{eq:YBC1_sphave}--\eqref{eq:YBC3_sphave} into the 6-point function    in Eq.~\eqref{eq:BB_expectation} to get the beat coupling contributions to the bispectrum covariance
\begin{align}
  \label{eq:CB_BC1}
  C_{\rm BC1} & = \frac{4624  }{ 441 } \sigma_W^2  B_{\rm m} B_{\rm m}' ,    \\
  \label{eq:CB_BC2}
  C_{\rm BC2} & =  \frac{ 68 }{1323  } \sigma_W^2 \bigg[   277  B_{\rm m} B_{\rm m}'  
    +  \frac{5}{2} \Big(  B_{\rm m} B_{\theta}'  +  B_{\rm m}' B_\theta    \Big)     \bigg] , \\
    \label{eq:CB_BC3}
  C_{\rm BC3} & =   \frac{1}{15876} \sigma_W^2 ( 277 B_{\rm m} + 5 B_\theta) ( 277 B_{\rm m}' + 5 B_\theta')  . 
\end{align}

In Fig.~\ref{fig:CB_HRS_SSC_ratio_11}, we compare the covariance obtained using Eq.~\eqref{eq:CB_BC1}-\eqref{eq:CB_BC3} (denoted as HRS for \cite{Hamilton:2005dx}) and the response approach described in the main text [Eqs.~\eqref{eq:CB_SSC} and \eqref{eq:Bk_response_SPT_global}]. Note that in this comparison we have used the global mean and equilateral triangle configuration. We find that the HRS method overestimates the covariance compared to the response results. The results are qualitatively similar to what we found for the case of the power spectrum.     

\begin{widetext}
  
\section{Supersample covariance contribution due to the large-scale tides }
\label{sec:SSC_tides}

Besides the long wavelength density perturbations, the large-scale tidal perturbation can also leave its imprint on small scales. In this Appendix we generalize the computations in the main text to include the effects of the tides on the small-scale density.  The large-scale tides also generate effects similar to the redshift space distortion \cite{Akitsu:2016leq,Akitsu:2017syq,Li:2017qgh}.  The power spectrum supersample covariance  due to the tides was investigated in \cite{Akitsu:2016leq}.  In this Appendix, we focus on the bispectrum supersample covariance due to the tides. 

\subsection{Modulated density in SPT}
Generalizing the calculations in Sec.~\ref{sec:response_SPT_longshortcoupling}, we compute the small-scale density modulated by the long wavelength perturbations including both density and tide. We preform the computations for the second and third order in SPT.  

\subsubsection{ Second order  }
From the second order SPT, we have
\beq
\delta_{\rm ls}^{(2)}( \bm{k} ) = 2 \int_{\bm{q}_0} \frac{ d^3 q }{ (2\pi)^3 }  F_2( \bm{q} , \bm{k}-\bm{q} ) \delta^{(1)}( \bm{q} ) \delta^{(1)}( \bm{k} - \bm{q} ), 
\eeq
where $\bm{q}_0$ denotes the Fourier mode of the long wavelength perturbation.  Expanding the $F_2( \bm{q} , \bm{k}-\bm{q} ) $ kernel and the $ \delta^{(1)}( \bm{k} - \bm{q} )$ about the long mode $\bm{q}$, we have
\begin{align}
  \delta_{\rm ls}^{(2)}( \bm{k} ) & = 2 \int_{\bm{q}_0} \frac{d^3 q }{ (2 \pi)^3 } \Big[  \frac{ \bm{q} \cdot \bm{k}  }{ 2 q^2  } + \frac{3}{14} + \frac{2}{7} \frac{ (\bm{k} \cdot \bm{q})^2  }{k^2 q^2}   \Big] \delta^{(1)} (\bm{q} ) [ \delta^{(1)} (\bm{k} )- \bm{q} \cdot \partial_{\bm{k}} \delta^{(1)}(\bm{k}  ) ] \nn \\
  &=  2 \int_{\bm{q}_0}\frac{ d^3 q }{ (2\pi)^3 }  \bigg[  \frac{\bm{q} \cdot \bm{k} }{ 2 q^2 } \delta^{(1)}(\bm{q}) \delta^{(1)}(\bm{k} ) + \Big(  \frac{ 3 }{14} + \frac{2}{ 7 } \frac{ ( \bm{k} \cdot\bm{q})^2 }{ k^2 q^2 }    \Big)  \delta^{(1)}( \bm{q} ) \delta^{(1)}(\bm{k})  - \frac{\bm{q} \cdot \bm{k} }{ q^2 } \bm{q} \cdot \partial_{ \bm{k} } \delta^{(1)}(\bm{k} ) \delta^{(1)}(\bm{q} )        \bigg]
\end{align}
The expansion of $F_2$ in the limit of $\bm{q} \rightarrow 0$  is related to the squeezed limit of the bispectrum. The lowest order term ($\propto q^{-1}$) is very general and can be understood as coming from the equivalence principle \cite{Peloso:2013zw,Kehagias:2013yd,Creminelli:2013mca}  and the rest are related to the dynamics. Unlike in the main text, we perform the computation by keeping the asymmetry explicit without spherically averaging over the angle of $ \hat{\bm{q}} $. Then we end up with
\beq
\label{eq:delta_2_density_tide}
\delta_{\rm ls}^{(2)}( \bm{k} ) = \bm{k}_i \lfloor \bm{q}_i \phi  \rfloor \delta^{(1)}(\bm{k})  + \frac{ 13}{21}  \lfloor \delta  \rfloor  \delta^{(1)}(\bm{k}) - \frac{1}{3} \bm{k}_i \partial_{\bm{k}_i} \delta^{(1)}(\bm{k}) \lfloor \delta  \rfloor  + \Big[  \frac{4}{7}\frac{  \bm{k}_i\bm{k}_j  }{ k^2 }  -  \bm{k}_i \partial_{\bm{k}_j } \Big] \delta^{(1)}(\bm{k}) \lfloor  \tau_{ij} \rfloor,   
\eeq
where the floor bracket denotes 
\beq
\lfloor \cdots \rfloor = \int_{\bm{q}_0} \frac{d^3 q  }{(2 \pi)^3 } \cdots , 
\eeq
and $\tau_{ij} $ is the tidal tensor 
\beq
\tau_{ij}( \bm{q} )  =  T_{ij}(\bm{q})  q^2 \phi \equiv \Big[  \frac{ \bm{q}_i \bm{q}_j  }{ q^2 } - \frac{1 }{ 3  } \delta_{ij} \Big]  q^2 \phi(\bm{q})  ,
\eeq
with $ \phi $ being the large-scale gravitational potential. The lowest order term in Eq.~\eqref{eq:delta_2_density_tide} (the dipole term in $\phi$) results from the large scale structure consistency relation, and it does not contribute to the physical polyspectrum.  The second and third terms are the same as those obtained by spherical averaging [Eq.~\eqref{eq:density2_modulated_SphAve}].  
The last term is the tidal part, and it is of the same order as the density terms.

\subsubsection{ Third order }
The modulated density from the third order coupling is 
\begin{align}
  \label{eq:delta_3order_modulated}
\delta_{\rm lss}^{(3)}( \bm{k} ) =   & 3 \int \frac{ d^3k_1 }{(2\pi)^3}\int \frac{ d^3k_2' }{(2\pi)^3} \int_{\bm{q}_0}  \frac{ d^3q }{(2\pi)^3}  (2 \pi)^3 \Ddel( \bm{k} - \bm{k}_1 - \bm{k}_2' - \bm{q} ) F_3(  \bm{k}_1, \bm{k}_2' , \bm{q} ) \delta^{(1)}(\bm{k}_1 ) \delta^{(1)}(\bm{k}_2' ) \delta^{(1)}(\bm{q} )  \nn \\
  \approx &  3  \int \frac{ d^3k_1 }{(2\pi)^3 }  \int_{\bm{q}_0}  \frac{ d^3q }{(2\pi)^3} \Big[ \frac{\bm{k} \cdot \bm{q} }{ q^2  } \frac{F_2( \bm{k}_1, \bm{k}_2  )  }{ 3 } + E_0 (  \bm{k}_1, \bm{k}_2  ) + \frac{ \bm{q}_i\bm{q}_j }{ q^2} E_{2}^{ij} (  \bm{k}_1, \bm{k}_2  )     \Big] \nn \\
  & \quad \quad \quad \quad \quad   \times \delta^{(1)}( \bm{k}_1 ) \delta^{(1)}( \bm{q} ) [  \delta^{(1)}( \bm{k}_2 )  - \bm{q} \cdot \partial_{ \bm{k} } \delta^{(1)}(\bm{k}_2 )   ] \nn \\
  = &  \bm{ k}_i \delta_{F_2} \lfloor \bm{q}_i \phi \rfloor   +   3 \delta_{E_0}\lfloor \delta \rfloor    + 3   \delta_{E_{2}}^{ij}   \lfloor \bm{q}_i \bm{q}_j \phi \rfloor  -  \bm{k}_i [ \partial_{\bm{k}_j} \delta_{F_2}   -   \delta_{\partial F_2 }^{j} ]  \lfloor \bm{q}_i \bm{q}_j \phi \rfloor  \nn \\
  = &  \bm{k}_i \delta_{F_2} \lfloor \bm{q}_i \phi \rfloor  +  \Big[  \frac{151}{126} \delta_{ F_2 } + \frac{5}{126} \delta_{ G_2 }     - \frac{ \bm{k}_i }{3} \partial_{\bm{k}_i } \delta_{F_2} \Big] \lfloor  \delta  \rfloor  
  +   (  \delta_{ H}^{ij}    - \bm{k}_i  \partial_{\bm{k}_j } \delta_{F_2  }  )  \lfloor \tau_{ij}  \rfloor    , 
\end{align}
where in the second line we have defined $ \bm{k}_2 \equiv \bm{k} -  \bm{k}_1 $. The second order density $ \delta_X^{i \cdots}$ is defined as
\beq
\delta_X^{i \cdots}(\bm{k}) = \int \frac{d^3 k_1 }{ (2 \pi)^3 } \int \frac{d^3 k_2 }{ (2 \pi)^3 } (2\pi)^3 \Ddel( \bm{k} - \bm{k}_1 - \bm{k}_2 ) X_{i \cdots}( \bm{k}_1, \bm{k}_2) \delta^{(1)}(\bm{k}_1 ) \delta^{(1)}(\bm{k}_2 ) .  
\eeq
The coupling kernels are given by
\begin{align}
  \partial F_2( \bm{k}_1, \bm{k}_2 ) & = \bm{k}\bigg[ \frac{1}{4} \Big( \frac{1}{k_1^2} + \frac{1}{k_2^2} \Big)  + \frac{2}{7}  \frac{ \bm{k}_1 \cdot \bm{k}_2 }{k_1^2 k_2^2 }  \bigg]   -  \frac{\bm{k}_1 \cdot \bm{k}_2 }{ 2 } \Big(  \frac{ \bm{k}_1 }{ k_1^4 } +  \frac{ \bm{k}_2 }{ k_2^4 }    \Big)  - \frac{2}{7} \frac{(\bm{k}_1 \cdot \bm{k}_2 )^2 }{ k_1^2 k_2^2  }\Big(   \frac{ \bm{k}_1 }{ k_1^2 } +  \frac{ \bm{k}_2 }{ k_2^2 }      \Big),  \\
  E_0 ( \bm{k}_1,\bm{k}_2) & =  \frac{ 5 (\bm{k}_1 \cdot \bm{k}_2)^2 + 9 \bm{k}_1 \cdot \bm{k}_2 (k_1^2 + k_2^2 ) + 13 k_1^2 k_2^2    }{ 63 k_1^2 k_2^2  }, \\
  E_2^{ii} ( \bm{k}_1,\bm{k}_2) & = \frac{1}{252 k_1^4 k_2^4 }\big[ 42 (\bm{k}_1 \cdot \bm{k}_2 )^2 ( k_1^4 + k_2^4)   +  24 (\bm{k}_1 \cdot \bm{k}_2 )^3 ( k_1^2 + k_2^2) - 21 k_1^2 k_2^2 (k_1^4 + k_2^4  )  \nn \\
   &  + 24 k_1^2 k_2^2 (\bm{k}_1 \cdot \bm{k}_2 ) (k_1^2 + k_2^2 ) + 32 k_1^2 k_2^2 ( \bm{k}_1 \cdot \bm{k}_2 )^2 + 22 k_1^4 k_2^4    \big],   \\
H_{ij}(\bm{k}_1,\bm{k}_2 ) & =  \frac{ \bm{k}_{1i}  \bm{k}_{1j}}{k_1^2}  f ( \bm{k}_1,\bm{k}_2 )  + 2  \frac{ \bm{k}_{1i} \bm{k}_{2j}}{k_1 k_2}  g ( \bm{k}_1,\bm{k}_2 )  +  \frac{  \bm{k}_{2i} \bm{k}_{2j}}{k_2^2 }  f ( \bm{k}_2,\bm{k}_1 ) ,   \\
f ( \bm{k}_1,\bm{k}_2 ) &= \frac{ 1}{ 42 k_1^2 k_2^2 k^2} \Big[  12 k_1^4 \bm{k}_1 \cdot \bm{k}_2 + 16 k_1^2   ( \bm{k}_1 \cdot \bm{k}_2 )^2  -   8  ( \bm{k}_1 \cdot \bm{k}_2 )^3 + 18 k_1^4 k_2^2   \nn \\
  & + 31 k_1^2 k_2^2 \bm{k}_1 \cdot \bm{k}_2  -  22 k_2^2  (\bm{k}_1 \cdot \bm{k}_2 )^2  + 14 k_1^2 k_2^4 -  9 k_2^4  \bm{k}_1 \cdot \bm{k}_2  \Big] ,    \\
g ( \bm{k}_1,\bm{k}_2 ) &= \frac{ 1 }{ 84 k_1 k_2 k^2 } \Big[ 21 (k_1^4 + k_2^4)  + 48 (\bm{k}_1 \cdot \bm{k}_2 )^2  + 70 (k_1^2 + k_2^2 ) \bm{k}_1 \cdot \bm{k}_2   + 50 k_1^2 k_2^2   \Big],   
\end{align}
where $\bm{k}$ denotes $\bm{k}_{1} + \bm{k}_2$.  Note that $f$ is not symmetric about its arguments while $g$ is. 

The expansion of $F_3$ in the low $q$ limit is a consequence of the consistency relation for the 4-point function.  In the last line of Eq.~\eqref{eq:delta_3order_modulated}, the dipole term  again does not contribute to the polyspectrum. The density terms are the same as those derived in Eq.~\eqref{eq:delta2s_long}.   The density part due to the enhancement in growth is simplified using 
\beq
3 \delta_{E_0} ( \bm{k})   + \delta_{E_{2}}^{ii} ( \bm{k})   + \frac{\bm{k}_i}{3} \delta_{\partial F_2 }^i  ( \bm{k})  = \frac{151}{126} \delta_{ F_2 } ( \bm{k})  + \frac{5}{126} \delta_{ G_2 } ( \bm{k}) . 
\eeq
The last part in Eq.~\eqref{eq:delta_3order_modulated} is the tidal contribution.  In the tidal part, $\delta_H^{ij} $ denotes
\beq                                                                 
\delta_H^{ij} ( \bm{k}) =   3  \delta_{E_{2}}^{ij}  ( \bm{k} )  +  \bm{k}_i  \delta_{\partial F_2 }^{j}  (\bm{k}), 
 \eeq
These terms share the same structure as those in Eq.~\eqref{eq:delta_2_density_tide}.


Recall that the factor of $-1/3 $ in the dilation term for the density  originates from the scaling $1/a^3$ for the dark matter [see the discussion around Eq.~\eqref{eq:density_pert_a3scaling}].   On the other hand, in the tidal part, we get the derivative term  $ - \bm{k}_i  \partial_{\bm{k}_j } \delta  $. From the prefactor $-1$,  we reason that the tidal perturbation must scale as
\beq
\frac{1 + \lfloor \tau_{ij} \rfloor }{a  } = \frac{ 1 }{ a_W }
\eeq
[cf.~Eq.~\eqref{eq:density_pert_a3scaling}].  In the principal frame, the tidal tensor is diagonal.  Thus we can absorb the tidal perturbations by introducing one scale factor along each principal axis.  For the tidal term, the global wave number $\bm{k}$ is related to the local one $\bm{k}_W $ as 
\beq
\bm{k}_{W i} = ( 1+ \lfloor \lambda_{i} \rfloor  )^{ - 1 } \bm{k}_i,  
\eeq
where $ \lambda_{i} $ is the eigenvalue of the tidal tensor.  
See similar discussions in  \cite{Hui:1995bw, BondMeyers1996,Ip:2016jji,Akitsu:2016leq}.

To summarize, up to first order in the long mode and second order in the short one, the modulated density reads
\begin{align}
  \label{eq:delta_modulation_density_tide}
  \delta( \bm{k} | \lfloor \delta \rfloor, \lfloor \tau \rfloor ) &  = \bm{k}_i (\delta^{(1)} + \delta_{F_2} ) \lfloor \bm{q}_i \phi  \rfloor   + \Big[ \frac{13}{21} \delta^{(1)}  +  \Big( \frac{ 151 }{126 } \delta_{ F_2}  +  \frac{ 5 }{126 } \delta_{ G_2}   \Big)  \Big]  \lfloor \delta \rfloor - \frac{1}{3} \bm{k}_i \partial_{ \bm k_i } (  \delta^{(1)} + \delta_{F_2} )    \lfloor \delta \rfloor \nn \\
 &  + \Big(  \frac{ 4}{7 }   \frac{ \bm{k}_i \bm{k}_j   }{ k^2 } \delta^{(1)} + \delta_H^{ij}   \Big)    \lfloor \tau_{ij} \rfloor  - \bm{k}_i \partial_{\bm{k}_j } (  \delta^{(1)} + \delta_{F_2}  )   \lfloor \tau_{ij} \rfloor. 
\end{align}
This is the generalization of Eq.~\eqref{eq:delta2s_long}.

\subsection{ The modulated bispectrum and response function } 

In this section, we compute the matter bispectrum with the long wavelength modulations.  We compute tree level bispectrum up to first order in long wavelength perturbation. 

It is easy to show that the bispectrum due to the gradient term  $ \lfloor \bm{q}  \phi   \rfloor  $ is  proportional to $\bm{k}_{123} B_{\rm M}(k_1,k_2,k_3) \Ddel( \bm{k}_{123} )  $. This shows that the gradient term indeed does not contribute to the bispectrum. 

The bispectrum arising from the density perturbation reads 
\beq
B_\delta(k_1,k_2,k_3) =  \bigg[ \frac{ 433 }{ 126 } B_{\rm M} (k_1,k_2,k_3)  +   \frac{5 }{126} B_{G_2} (k_1,k_2,k_3)    -    \frac{1}{ 3}  \sum_{i=1}^3 \frac{ \partial }{\partial \ln k_i }   B_{\rm M} (k_1,k_2,k_3)   \bigg]  \lfloor \delta \rfloor  .
\eeq


The bispectrum due to the growth part of the tidal tensor is given by
\begin{align}
  B_\tau^{(1)} (k_1,k_2,k_3) = &  \frac{4}{ 7 } \lfloor \tau_{ij}  \rfloor  \Big( \frac{ \bm{k}_{1i} \bm{k}_{1j}   }{k_1^2} +  \frac{ \bm{k}_{2i} \bm{k}_{2j}   }{k_2^2}  \Big)  2 F_2 (\bm{k}_1,\bm{k}_2 ) P(k_1) P(k_2) + 2 \cyc    \nn \\
& +  2   \lfloor \tau_{ij}  \rfloor  H_{ij} ( \bm{k}_1,  \bm{k}_2 ) P(k_1) P(k_2)  + 2 \cyc 
\end{align}
To compute the contribution due to the dilation part of the tidal term, we note that
\begin{align}
  \delta( \bm{k} ) - \lfloor \tau_{ij} \rfloor  \bm{k}_j \partial_{ \bm{k}_i  } \delta( \bm{k} )  \approx      \delta ( \bm{k}_i - \lfloor \tau_{ij} \rfloor  \bm{k}_j )   =   \delta \big( ( \delta^{\rm K}_{ij} - \lfloor \tau_{ij} \rfloor )  \bm{k}_j \big)  .    
\end{align}
By considering
\begin{align}
   &  \Big\langle  \delta^{(1)}\big( ( \delta^{\rm K}_{ij} - \lfloor \tau_{ij} \rfloor )  \bm{k}_{1j} \big)  \delta^{(1)}\big(  (\delta^{\rm K}_{ij} - \lfloor \tau_{ij} \rfloor )  \bm{k}_{2j} \big)      \delta^{(1)}\big( ( \delta^{\rm K}_{ij} - \lfloor \tau_{ij} \rfloor )  \bm{k}_{3j} \big)  \Big\rangle  + 2 \cyc    \nn \\
 \approx  &  (2 \pi)^3    \Ddel(\bm{k}_{123}) \bigg[  B_{\rm M}(k_1,k_2,k_3) - \sum_{a=1}^3 \frac{ \bm{k}_{ai} \bm{k}_{aj}}{ k_a^2  }    \lfloor \tau_{ij  } \rfloor \frac{ \partial   }{ \partial \ln k_a } B_{\rm M}(k_1,k_2,k_3)\bigg],
\end{align}
we get the dilation contribution to the tidal part
\beq
  B_\tau^{(2)} (k_1,k_2,k_3) =  -   \lfloor \tau_{ij  } \rfloor   \sum_{a=1}^3 \frac{ \bm{k}_{ai} \bm{k}_{aj}}{ k_a^2  }   \frac{ \partial   }{ \partial \ln k_a } B_{\rm M}(k_1,k_2,k_3) .
\eeq

Before computing the response function, we take the long-wavelength density and tide perturbation to be
\beq
\label{eq:delta_tau_perturbation_form}
\delta_{\rm l}( \bm{q} ) =  (2 \pi)^3 \Ddel( \bm{q}- \bm{q_0} ) \delta_{\rm b},   \qquad   \tau_{\rm l}^{ij} ( \bm{q} ) =  (2 \pi)^3 \Ddel( \bm{q}- \bm{q_0} ) \tau_{\rm b}^{ij}, 
\eeq
then we have  $ \lfloor \delta \rfloor = \delta_{\rm b} $ and  $  \lfloor \tau_{ij} \rfloor =  \tau_{ \mathrm{b} }^{  ij} $.    
The response functions are given by
\begin{align}
  \label{eq:response_density}
  \frac{\partial  }{ \partial \delta_{\rm b}} B_{\delta}( k_1,k_2,k_3)  \Big|_{ \delta_{\rm b} = 0 } & =  \frac{ 433 }{126} B_{\rm M} (k_1,k_2,k_3)  +  \frac{5 }{126} B_{G_2} (k_1,k_2,k_3)   - \frac{1}{ 3}  \sum_{i=1}^3 \frac{ \partial }{\partial \ln k_i }   B_{\rm M} (k_1,k_2,k_3) ,  \\
    \label{eq:response_tide}
\frac{\partial  }{ \partial \tau_{\mathrm{ b} }^{ ij }} B_{\tau}( \bm{k}_1,\bm{k}_2,\bm{k}_3)  \Big|_{ \tau_{\mathrm{ b} ij } = 0 } & =  \bigg[    \frac{4}{ 7 } \Big( \frac{ \bm{k}_{1i} \bm{k}_{1j}   }{k_1^2} +  \frac{ \bm{k}_{2i} \bm{k}_{2j}   }{k_2^2}  \Big)  2 F_2 (\bm{k}_1,\bm{k}_2 ) + 2  H_{ij}( \bm{k}_1,\bm{k}_2 )  \bigg]    P(k_1) P(k_2) + 2 \cyc    \nn \\
&  \qquad \qquad - \sum_{a=1}^3 \frac{ \bm{k}_{ai} \bm{k}_{aj}}{ k_a^2  }    \frac{ \partial   }{ \partial \ln k_a } B_{\rm M}(k_1,k_2,k_3) .    
\end{align}
The response function to the long wavelength density and to the tide are of the same order of magnitude. However, the density response function only depends on the shape of the triangle, while the response to the tide also depends on the orientation of the triangle.  This will be important for the final numerical values of the covariance later on.

\subsection{ Supersample covariance with tides }

Now we generalize the computations of the supersample covariance in Sec.~\ref{sec:SSC_Bk_derivation} to include the effect of the tides:
\begin{align}
  \label{eq:6point_longmode_tides_expand1} 
&   \Big\langle  \delta(\bm{p}_1)   \delta(\bm{p}_2)   \delta(\bm{p}_3 - \bm{q}_{123} )    \delta(\bm{p}_1')  \delta(\bm{p}_2')  \delta(\bm{p}_3' - \bm{q}_{123}' )    \Big\rangle   \nn \\
  \approx  & \Big\langle  \Big[ \langle  \delta(\bm{p}_1)   \delta(\bm{p}_2)   \delta(\bm{p}_3) \rangle
    + \delta_{\rm l}( - \bm{q}_{123} ) \frac{ \partial  }{ \partial \delta_{\rm l}( \bm{q} )   }   \langle  \delta(\bm{p}_1)   \delta(\bm{p}_2)   \delta(\bm{p}_3 + \bm{q}) \rangle \Big|_{\delta_{\rm l} =0 }   +
 \tau_{\mathrm{l}}^{ ij}(  - \bm{q}_{123} )   \frac{ \partial  }{ \partial \tau_{\mathrm{l}}^{ ij}( \bm{q} )   }   \langle  \delta(\bm{p}_1)   \delta(\bm{p}_2)   \delta(\bm{p}_3 + \bm{q}) \rangle \Big|_{\tau_{\rm l} = 0 } \Big]    \times    \nn \\
   \Big[   \langle &  \delta(\bm{p}_1')   \delta(\bm{p}_2')   \delta(\bm{p}_3') \rangle  + \delta_{\rm l} (- \bm{q}_{123}' ) \frac{ \partial  }{ \partial \delta_{\rm l} (\bm{q}')  } \langle  \delta(\bm{p}_1')   \delta(\bm{p}_2')   \delta(\bm{p}_3' + \bm{q}' ) \rangle \Big|_{\delta_{\rm l} =0 }    + \tau_{ \mathrm{l}}^{  mn}(  - \bm{q}_{123}' )   \frac{ \partial  }{ \partial \tau_{\mathrm{l}}^{  mn}( \bm{q}' )   }   \langle  \delta(\bm{p}_1')   \delta(\bm{p}_2')   \delta(\bm{p}_3' + \bm{q}') \rangle \Big|_{\tau_{\rm l} = 0 } \Big]  \Big\rangle_{ \delta_{\rm l}, \tau_{\rm l} }   \nn \\
 \approx  &   \langle  \delta(\bm{p}_1)   \delta(\bm{p}_2)   \delta(\bm{p}_3 ) \rangle  \langle \delta(\bm{p}_1')  \delta(\bm{p}_2')  \delta(\bm{p}_3' )   \rangle   + (2 \pi)^3 \Ddel(\bm{q}_{123} + \bm{q}'_{123} ) P_{\rm l} (q_{123} )    \frac{\partial}{\partial \delta_{\rm b} } B(p_1,p_2,p_3 | \delta_{\rm b} )   \Big|_{\delta_{\rm b} = 0 }    \frac{\partial}{\partial \delta_{\rm b} }   B(p_1',p_2',p_3'|\delta_{\rm b})  \Big|_{\delta_{\rm b} = 0 }  \nn \\
 & +  ( 2 \pi )^3 T_{ij}( \bm{q}_{123}   )T_{mn}( \bm{q}_{123}) \Ddel( \bm{q}_{123} + \bm{q}_{123}' )   P_{\rm l}(q_{123})   \frac{\partial}{\partial \tau_{\mathrm{ b}}^{ ij } } B(\bm{p}_1,\bm{p}_2,\bm{p}_3 | \tau_{\mathrm{ b}}^{ ij } )   \Big|_{\tau_{\rm b} = 0 }    \frac{\partial}{\partial \tau_{\mathrm{ b}}^{ mn} }   B(\bm{p}_1',\bm{p}_2',\bm{p}_3' | \tau_{\mathrm{ b}}^{ mn}  )  \Big|_{\tau_{\mathrm{ b} } = 0 }   
\end{align}
As in Ref.~\cite{Akitsu:2016leq}, we have neglected the cross terms between the density and the tidal field ($\propto \langle \delta_{\rm l}  \tau_{\rm l}^{ij} \rangle$). We will justify this shortly.  In deriving response function in Eq.~\eqref{eq:6point_longmode_tides_expand1}, we have taken the density and tide perturbations of the form Eq.~\eqref{eq:delta_tau_perturbation_form}.  In Eq.~\eqref{eq:6point_longmode_tides_expand1}, the first term  is canceled by $\langle \hat{B} \hat{B}' \rangle $. The second term is the supersample covariance due to the long wavelength density, while the last one is due to the tidal perturbations.

The integral over the window functions can be simplified as in Eq.~\eqref{eq:Wconvolution_intg}. The net results are
\begin{align}
S      & =  \int \frac{ d^3 q }{(2\pi)^3 } \bigg[ \frac{ W (\bm{q}) }{V} \bigg]^2P_{\rm l} (q),  \\
S_{ij}  & =  \int \frac{ d^3 q }{(2\pi)^3 } \bigg[ \frac{ W (\bm{q}) }{V} \bigg]^2 T_{ij}(\bm{q})  P_{\rm l} (q) ,  \\
\label{eq:Sijmn}
S_{ijmn}& =   \int \frac{ d^3 q }{(2\pi)^3 } \bigg[ \frac{ W (\bm{q}) }{V} \bigg]^2 T_{ij}(\bm{q}) T_{mn}(\bm{q})  P_{\rm l} (q).  
\end{align}
Note that $S$ is $\sigma_W^2$ in the main text.  
For the spherically symmetric window we have  $S_{ij} = 0 $. Thus for a reasonable window we anticipate that $ S_{ij} \ll S $.  That is why we have neglected the cross term in Eq.~\eqref{eq:6point_longmode_tides_expand1}.   For $S_{ijmn}$ only the following elements are non-vanishing if the window is spherically symmetric  ($i \ne j $) 
\begin{align}
  S_{iiii} = \frac{ 4 }{ 45 } S, \quad   S_{iijj} = - \frac{ 2 }{ 45 } S, \quad   S_{ijij} = S_{ijji} = \frac{ 1 }{ 15 } S .
\end{align}

Finally we arrive at the supersample covariance due to the density and tide
\begin{align}
  \label{eq:SSC_density_appendix}
  C_{\rm SSC}^{\delta}( p_1,p_2,p_3; p_1',p_2',p_3'  )  &= S   \,  \frac{\partial}{\partial \delta_{\rm b} } B(p_1,p_2,p_3 | \delta_{\rm b} )   \Big|_{\delta_{\rm b} = 0 }    \frac{\partial}{\partial \delta_{\rm b} }   B(p_1',p_2',p_3'|\delta_{\rm b})  \Big|_{\delta_{\rm b} = 0 } ,  \\
  \label{eq:SSC_tide_appendix}
  C_{\rm SSC}^{\tau}( \bm{p}_1,\bm{p}_2,\bm{p}_3; \bm{p}_1',\bm{p}_2',\bm{p}_3'   )  &= S_{ij mn }      \frac{\partial}{\partial \tau_{\mathrm{ b}}^{ ij } } B(\bm{p}_1,\bm{p}_2,\bm{p}_3 | \tau_{\mathrm{ b}}^{ ij } )   \Big|_{\tau_{\mathrm{b}}^{ ij } = 0 }    \frac{\partial}{\partial \tau_{\mathrm{ b}}^{ mn} }   B(\bm{p}_1',\bm{p}_2',\bm{p}_3' | \tau_{\mathrm{ b}}^{ mn}  )  \Big|_{\tau_{\mathrm{ b} }^{ mn} = 0 } .
\end{align}

\end{widetext}

\subsection{ Numerical results } 

The bispectrum only depends on the shape of the triangle, and so it is estimated by averaging over all the triangles of the same shape but different orientations,
\beq
B(k_1,k_2,k_3) = \frac{ 1 }{ N_{k_1 k_2 k_3 } } \sum_{\alpha \in  \{ k_1 k_2 k_3 \} } B_\alpha( \bm{k}_1, \bm{k}_2, \bm{k}_3 ),
\eeq
where $  N_{k_1 k_2 k_3 }  $ denotes the number of possible orientations of the configurations $k_1k_2k_3$ in the summation. In the covariance we consider
\begin{align}
      & \langle  B(k_1,k_2,k_3) B(k_1',k_2',k_3') \rangle  
  =   \frac{ 1 }{ N_{k_1 k_2 k_3 } }  \frac{ 1 }{ N_{k_1' k_2' k_3' } }   \nn \\
  & \times \sum_{\alpha \in  \{ k_1 k_2 k_3 \} }  \sum_{\beta \in  \{ k_1' k_2' k_3' \} }  \langle   B_\alpha( \bm{k}_1, \bm{k}_2, \bm{k}_3 )  B_\beta( \bm{k}_1', \bm{k}_2', \bm{k}_3' ) \rangle.    
\end{align}
The covariance contribution to $\langle B B' \rangle $ from the tide reads
\begin{align}
   \frac{ 1 }{ N_{k_1 k_2 k_3 } }   \sum_{\alpha \in  \{ k_1 k_2 k_3 \} }    \bar{C}_{\rm SSC }^{\tau} ( \bm{k}_1,\bm{k}_2,\bm{k}_3,  k_1',k_2',k_3'), 
\end{align}
with $ \bar{C }_{\rm SSC }^{\tau} $ given by 
\begin{align}
\label{eq:SSC_tide_estimator_average}
& \bar{C}_{\rm SSC }^{\tau} ( \bm{k}_1,\bm{k}_2,\bm{k}_3, k_1',k_2',k_3') \nn \\
= &  \frac{ 1 }{ N_{k_1' k_2' k_3' } }
  \sum_{\beta \in  \{ k_1' k_2' k_3' \} }   C_{\rm SSC }^{\tau} ( \bm{k}_1,\bm{k}_2,\bm{k}_3,  \bm{k}_1',\bm{k}_2',\bm{k}_3').
\end{align}
Because the final result only depends on the relative orientation between the triplet $\bm{k}_1 \bm{k}_2 \bm{k}_3  $ and $\bm{k}_1' \bm{k}_2' \bm{k}_3' $,  we only need to focus on $ \bar{C}_{\rm SSC }^{\tau} $. We compute it using Monte Carlo methods as follows: after picking a fixed triplet for  $\bm{k}_1 \bm{k}_2 \bm{k}_3$ we randomly rotate the triangle $\bm{k}_1'\bm{k}_2'\bm{k}_3'$ to sample the sum.

Although the magnitude of the tide response Eq.~\eqref{eq:response_tide} and  $S_{ijmn} $ are comparable to their density counterparts,  because of the averaging in Eq.~\eqref{eq:SSC_tide_estimator_average} and the fact that the tide response function is anisotropic, after spatial averaging, the value of $\bar{C}_{\rm SSC}^\tau$ is substantially reduced. For example if we estimate $\bar{C}_{\rm SSC }^{\tau}$ using the equilateral triangle configuration of  $k = 0.14 \MpcOh $, the tidal contribution is about three orders of magnitude smaller than the density one. Because the reduction stems from the averaging of the anisotropic response function, the supersample covariance due to the tide is expected to be small relative to the density one for generic window functions.  We comment that the bispectrum becomes anisotropic in the redshift space, and this importance of the tidal term could be potentially larger in that case. We leave the investigation in redshift space for future work.



\bibliography{references}

\end{document}